# Modular SIMD arithmetic in Mathemagix

Joris van der Hoeven[a], Grégoire Lecerf[b]

Laboratoire d'informatique LIX, UMR 7161 CNRS
Campus de l'École polytechnique
Route de Saclay
91128 Palaiseau Cedex, France

a. *Email:* vdhoeven@lix.polytechnique.fr
b. *Email:* lecerf@lix.polytechnique.fr

Guillaume Quintin

Laboratoire XLIM, UMR 7252 CNRS
Université de Limoges
123, avenue Albert Thomas
87060 Limoges Cedex, France

*Email:* guillaume.quintin@unilim.fr

*Version of June 29, 2014*

Modular integer arithmetic occurs in many algorithms for computer algebra, cryptography, and error correcting codes. Although recent microprocessors typically offer a wide range of highly optimized arithmetic functions, modular integer operations still require dedicated implementations. In this article, we survey existing algorithms for modular integer arithmetic, and present detailed vectorized counterparts. We also present several applications, such as fast modular Fourier transforms and multiplication of integer polynomials and matrices. The vectorized algorithms have been implemented in C++ inside the free computer algebra and analysis system Mathemagix. The performance of our implementation is illustrated by various benchmarks.

## 1. Introduction

**Motivations**

During the past decade, major manufacturers of microprocessors have changed their focus from ever increasing clock speeds to putting as many cores as possible on one chip and to lower power consumption. One approach followed by leading constructors such as Intel® and AMD® is to put as many x86 compatible processors on one chip. Another approach is to rely on new simplified processing units, which allows to increase the number of cores on each chip. Modern *Graphics Processing Units* (GPU) have been designed according to this strategy.

As powerful as multicore architectures may be, this technology also comes with its drawbacks. Besides the increased development cost of parallel algorithms, the main disadvantage is that the high degree of concurrency allowed by multicore architecture often constitutes an overkill. Indeed, many computationally intensive tasks ultimately boil down to classical mathematical building blocks such as matrix multiplication or *fast Fourier transforms* (FFTs).

In many cases, the implementation of such building blocks is better served by simpler parallel architectures, and more particularly by the *Single Instruction, Multiple Data* (SIMD) paradigm. A limited support for this paradigm has been integrated into the x86 processor families with the Intel® MMX™ and SSE (*Streaming SIMD Extensions*) technologies. More recently, the advent of Intel AVX (*Advanced Vector Extensions*) and AVX-512 has further empowered this model of computation. Historically speaking, vector computers have played an important role in "High Performance Computing" (HPC). An interesting question is whether more powerful SIMD instructions can also be beneficial to other areas.

More specifically, this paper concerns the efficiency of the SIMD approach for doing exact computations with basic mathematical objects such as integers, rational numbers, modular integers, polynomials, matrices, etc. Besides high performance, we are therefore interested in the reliability of the output. For this reason, we like to coin this area as *High Quality Computing* (HQC).

One important mathematical building block for HQC is fast modular arithmetic. For instance, products of multiple precision integers can be computed efficiently using FFTs over finite fields of the form $\mathbb{F}_p$ with $p = k\, 2^l + 1$, where $p$ typically fits within 32 or 64 bits, and $l$ is as large as possible under this constraint [55]. Similarly, integer matrices can be multiplied efficiently using such FFTs or Chinese remaindering.

The aim of this paper is twofold. First of all, we adapt known algorithms for modular integer arithmetic to the SIMD model. We carefully compare various possible implementations as a function of the bit-size of the modulus and the available SIMD instructions. In the second part of the paper, we apply our implementations to several fundamental algorithms for exact arithmetic, namely to the multiplication of integers, integer matrices, modular polynomials and modular polynomial matrices. We show that our implementations based on modular arithmetic outperform existing implementations in many cases, sometimes by one order of magnitude.

The descriptions of the algorithms are self-contained and we provide implementation details for the SSE and AVX instruction sets. All implementations were done in C++ and our software is freely distributed within the computer algebra and analysis system Mathemagix [41, 42, 44, 45, 51] (revision $\geqslant 9170$).

## Related work

There is large literature on modular integer arithmetic. The main operation to be optimized is multiplication modulo a fixed integer $p > 1$. This requires an efficient reduction of the product modulo $p$. The naive way of doing this reduction would be to compute the remainder of the division of the product by $p$. However, divisions tend to be more expensive than multiplications in current hardware. The classical solution to this problem is to pre-compute the inverse $p^{-1}$ so that all divisions by $p$ can be transformed into multiplications.

In order to compute the product $c$ of $a$ and $b$ modulo $p$, it suffices to compute the integer floor part $q$ of the rational number $a\,b\,p^{-1}$ and to deduce $c = a\,b - q\,p$. The most obvious algorithm is to compute $q$ using floating point arithmetic. However, this approach suffers from a few drawbacks. First, in order to ensure portability, the processor should comply to the IEEE-754 standard, which ensures that rounding errors are handled in a predictable way. Secondly, the size of the mantissa can be twice as large as the size of the modulus. Back and forth conversions from and to integers may then be required, which are expensive on some platforms. For more details we refer the reader to [22, Chapter 14] and [3, 4]. Many recent processors contain *fused multiply-add* instructions (FMA), which facilitate taking full advantage of floating point operations; see for instance [60, Chapitre 14].

*Barrett's algorithm* [7] provides an alternative to the floating point approach. The idea is to rescale the floating point inverse of $p^{-1}$ and to truncate it into an integer type. For some early comparisons between integer and floating point operations and a branch-free variant, we refer to [32]. This approach is also discussed in [23, Section 16.9] for particular processors. For multiple precision integers, algorithms were given in [6, 39].

Another alternative approach is to precompute the inverse $p^{-1}$ of the modulus $p$ as a 2-adic integer. This technique, which is essentially equivalent to *Montgomery's algorithm* [52], only uses integer operations, but requires $p$ to be odd. Furthermore, modular integers need to be encoded and decoded (with a cost similar to one modular product), which is not always convenient. Implementations have been discussed in [49]. A generalization to even moduli was proposed in [48]. It relies on computing separately modulo $2^k$ and $\tilde{p}$ (such that $\tilde{p}$ is odd, and $p = 2^k \tilde{p}$) *via* Chinese remaindering. Unfortunately, this leads to a significant overhead for small moduli. Comparisons between Barrett's and Montgomery's product were given in [15] for values of $k$ corresponding to a few machine words. A recent survey can be found in [54] with a view towards hardware implementation.

Modular arithmetic and several basic mathematical operations on polynomials and matrices have been implemented before on several types of parallel hardware (multicore architectures and GPUs); see for instance [8, 9, 29, 33]. Although the present article focuses on x86 compatible processor families and their recent extensions, the algorithms we describe are intended to be useful for all platforms supporting SIMD technology.

Our applications to matrix product over large polynomials or integers are central tasks to many algorithms in computer algebra [2, 12, 26, 50]. For a sample of recent targeted implementations, the reader might consult [30, 38, 53].

## Our contributions

One difficulty with modular arithmetic is that the most efficient algorithms are strongly dependent on the bit-size of the modulus $p$, as well as the availability and efficiency of specific instructions in the hardware. This is especially so in the case of SIMD processors, for which the precise instruction sets still tend to evolve rather quickly. In this paper, we have attempted to be as exhaustive as possible: we implemented all classical strategies, while hand optimizing all algorithms as a function of the available SIMD instructions.

The fast sequential algorithms for modular reduction from the previous section all involve some branching in order to counterbalance the effect of rounding. Our first contribution is to eliminate all branching so as to make vectorization possible and efficient. Our second contribution is a complete implementation of the various approaches as a function of the available SIMD instructions. For Barrett's and Montgomery's algorithms, we consider both the SSE 4.2 and AVX 2 instruction sets, for all types of supported integers. For the floating point approach, we have implementations with and without SSE 4.1, AVX, and FMA extensions. We finally provide detailed benchmarks for the performance of the different approaches in practice. The observed speedups with respect to the "scalar" versions (i.e. without vectorization) nearly reflect theoretical expectations.

High performance libraries such as GMP [31], MPIR [35], MPFR [24], FLINT [34] for computing with large integers, floating point numbers in arbitrary precision, polynomials and matrices are written in C and mostly aim at providing users with very fast mathematical operations. In fact, the internal representations and algorithms operating on them are not expected to be accessed nor replaced *a posteriori*. Instead, the C++ libraries of MATHEMAGIX provide users with high level interfaces to data structures and mathematical objects, but also with interfaces to internal representations and algorithms. Users can assemble algorithms for specific purposes, fine tune thresholds, and can easily replace an algorithm by another one *a posteriori*. In Section 4, we explain the main design principles. In particular, we show how our modular numbers are implemented and how our approach makes it simple to develop SIMD variants of algorithms which benefit from hardware vectorization features.

A third contribution of this article is the application of our modular arithmetic to an SIMD version of the fast Fourier transform algorithm in a prime field of the form $\mathbb{Z}/p\mathbb{Z}$, where $p-1$ is a multiple of a large power of two. Our implementations outperform existing software and we report on timings and comparisons for integer, polynomial and matrix products.

Besides low level software design considerations, our main research goals concern algorithms for solving polynomial systems, for effective complex analysis, and error correcting codes. MATHEMAGIX includes several of our recent algorithms [5, 10, 11, 43]. From our experience, a strict bottom-up approach to mathematical software design prevents users from implementing high level algorithms efficiently. In other words, assembling algorithms from a strict mathematical point of view does not always lead to the best performance. Interfaces to algorithms and the ability to reassemble them for higher level operations with a sufficient level of genericity turns out to be very useful. For instance multiplying integer matrices by performing integer products and additions in sequence according to the schoolbook algorithm can be improved if one has access to relatively generic implementations of the low level integer product sub-functions. This specific example is adressed in our last section.

## Conventions

Throughout this article, timings are measured on a platform equipped with an INTEL® CORE™ *i7-4770* CPU @ 3.40 GHz and 8 GB of *1600 MHz DDR3*. It runs the JESSIE GNU DEBIAN® operating system with a LINUX® kernel version 3.12 in 64 bit mode. Care has been taken for avoiding CPU throttling issues while measuring timings. Nevertheless timings often vary in a non negligible range sometimes over 20% and we thus measure average timings. We compile mainly with GCC [27] version 4.8.2. For pseudocode we use the operators of the C99 language with their exact meanings and the usual integer types from `inttypes.h`. Note also that integer promotions do not invalidate the algorithms and proofs in this paper. We refer the reader to the C99 (ISO/IEC 9899:1999) standard Sections 6.2.5 and 6.3.1.1 for further details [16]. For simplicity we assume that the operator `>>` always implements the *right arithmetic shift*.

## Brief survey of SIMD technology

For completeness, we conclude this introduction with recalling basic facts on SIMD technology. This technology can be seen as a type of parallelism where multiple processing units simultaneously execute the same instruction on multiple data. We can think of an SIMD processor as a single CPU which operates on registers which are really vectors of scalar data. The main basic characteristics of SIMD processors are the following:
- The scalar data types which are supported;

- The total bit-size $b$ of vector registers;
- The number $v$ of vector registers;
- For each scalar data type $T$, the instruction set for vector operations over $T$;
- The instruction set for other operations on SIMD registers, such as communication with the main memory or permutations of entries.

Modern SIMD processors typically support both floating point types (`float` and `double`) and integer types (8, 16, 32, and 64-bit integers). For a scalar type $T$ of bit-size $n$, we operate on vectors of $b/n$ coefficients in $T$.

Currently, the most common SIMD technologies are SSE (with $b=128$ and $v=16$), AVX (with $b=256$ and $v=16$) and AVX-512 (with $b=512$ and $v=32$). For instance, an AVX enabled processor can execute an operation on a vector of 16 integers of bit-size 16 in unit time. It should be noticed that these bit-sizes $b \leqslant 512$ are rather modest in comparison to their historical ancestors, such as the CDC STAR-100 and CRAY® vector computers (CDC was a trademark of CONTROL DATA CORPORATION). One advantage of shorter vectors is that it remains feasible to provide instructions for permuting the entries of vectors in quite general ways (i.e. "communications" remain relatively inexpensive, whenever needed).

In this paper, our SIMD algorithms are described using *compiler intrinsics*. Most of the time, the semantics of the SIMD types and instructions in this paper is quite straightforward. Let us give a few examples:

- The types `__m128i` and `__m128d` correspond to packed 128-bit integer and floating point vectors. They are supported by SSE enabled processors.
- The instruction `_mm_add_epi64` corresponds to the addition of two vectors of 64-bit signed integers of type `__m128i` (so the vectors are of length 2).
- The predicate `_mm_cmpgt_epi64` corresponds to a component-wise $>$ test on two vectors of 64-bit signed integers of type `__m128i`. The boolean results `true` and `false` are encoded by the signed integers `-1` and `0`.
- The instruction `_mm_mul_pd` corresponds to the multiplication of two vectors of double precision floating point numbers of type `__m128d` (so the vectors are of length 2).

We notice that all floating point arithmetic conforms to the IEEE-754 standard. In particular, results of floating point operations are obtained through correct rounding of their exact mathematical counterparts, where the global rounding mode can be set by the user. Some processors also provide fused multiply add (FMA) and subtract instructions `_mm_fmadd_pd` and `_mm_fmsub_pd` which are useful in Section 3. Another less obvious but useful instruction that is provided by some processors is:

- `_mm_blendv_pd` ($\vec{x}$, $\vec{y}$, $\vec{m}$) returns a vector $\vec{z}$ with $z_i = y_i$ whenever the most significant bit of $m_i$ is set and $z_i = x_i$ otherwise (for each $i$). For floating point numbers, it should be noticed that the most significant bit of $m_i$ corresponds to the sign bit.

For more details about SSE and AVX intrinsics we refer to the INTEL® intrinsic guides [46, 47], and also to [21] for useful comments and practical recommendations.

Let us mention that a standard way to benefit from SIMD technology is to rely on *auto-vectorization* features of compilers. Languages such as C and C++ have not yet been extended to support SIMD technology in a standard and more explicit manner. In particular, current SIMD types are not even *bona fide* C/C++ types. Programming *via* intrinsics contains a certain number of technical pitfalls. In particular streaming load and store instructions are slowed down whenever memory addresses are not aligned on the size of the vectors. It is therefore left to the programmer to wrap memory allocation functions or to rely on specific features of the compiler.

## 2. Modular operations *via* integer types

If $a$ and $b$ are nonnegative integers, then $a$ quo $b$ and $a$ rem $b$ represent the *quotient* and the *remainder* in the long division of $a$ by $b$ respectively. We let U denote an unsigned integer type of bit-size $n$ assumed to be at least 2 for convenience. This means that all the integers in the range from 0 to $2^n - 1$ can be represented in U by their binary expansion. The corresponding type for signed integers of bit-size $n$ is written I: all integers from $-2^{n-1}$ to $2^{n-1} - 1$ can be represented in I. The type L represents unsigned integers of size at least $2\,n$. Let $p$ be a nonnegative integer of bit-size at most $m$ with $m \leqslant n$. For efficiency reasons we consider that $n$ and $m$ are quantities known at compilation time, whereas the actual value of $p$ is only known at execution time. In fact, deciding which elementary modular arithmetic algorithm to use at runtime according to the bit-size of $p$ is of course too expensive. For convenience we identify the case $p = 0$ to computing modulo $2^n$. Modulo $p$ integers are stored as their representative in $\{0, ..., p-1\}$.

### 2.1. Modular sum

Although modular sum seems rather easy at first look, we quickly summarize the known techniques and pitfalls.

#### 2.1.1. Unvectorized implementations

Given $x$ and $y$ modulo $p$, in order to compute $(x + y) \operatorname{rem} p$, we can first compute $x + y$ in U and subtract $p$ when an overflow occurs or when $x + y \geqslant p$, as detailed in the following function:

**Function 1**

    **Input.** Integers $x$ and $y$ modulo $p$.
    **Output.** $(x + y) \operatorname{rem} p$.

```
U add_mod (U x, U y) {
 1. U a = x + y;
 2. if (a < x) return a - p;
 3. return (a >= p) ? a - p : a; }
```

Of course, when $m \leqslant n - 1$, no overflow occurs in the sum of line 1, and line 2 is useless. If branching is more expensive than shifting and if $m \leqslant n - 1$, then one can compute I a = x + y - p and return a + ((a >> (n-1)) & p), where we recall that >> implements the right arithmetic shift. It is important to program both versions and determine which approach is the fastest for a given processor. Negation and subtraction can be easily implemented in a similar manner.

#### 2.1.2. Implementations with SSE 4.2 and AVX 2

Arithmetic operations on packed integers are rather well supported by SSE 4.2, uniformly for various types of integers. Let $\vec{p}$ represent the packed $n$-bit integer of type \_\_m128i, whose entries are filled with $p$. In order to avoid branching in Function 1, one can compute U a = x + y and return min (a, a - p) when $m \leqslant n - 1$. This approach can be straightforwardly vectorized for packed integers of bit-sizes 8, 16 and 32, as exemplified in the following function:

**Function 2**

    **Input.** Packed 32-bit integers $\vec{x}$ and $\vec{y}$ modulo $\vec{p}$, assuming $m \leqslant n - 1$.
    **Output.** $(\vec{x} + \vec{y}) \operatorname{rem} \vec{p}$.

```
__m128i add_mod_1_epu32 (__m128i x⃗, __m128i y⃗) {
  1. __m128i a⃗ = _mm_add_epi32 (x⃗, y⃗);
  2. return _mm_min_epu32 (a⃗, _mm_sub_epi32 (a⃗, p⃗)); }
```

Since the min operation does not exist on packed 64-bit integers, we use the following function, where $\vec{0}$ represents the packed $n$-bit integer of type `__m128i` filled with 0:

**Function 3**

**Input.** Packed 64-bit integers $\vec{x}$ and $\vec{y}$ modulo $\vec{p}$, assuming $m \leqslant n-1$.
**Output.** $(\vec{x}+\vec{y}) \operatorname{rem} \vec{p}$.

```
__m128i add_mod_1_epu64 (__m128i x⃗, __m128i y⃗) {
  1. __m128i a⃗ = _mm_sub_epi64 (_mm_add_epi64 (x⃗, y⃗), p⃗);
  2. __m128i b⃗ = _mm_cmpgt_epi64 (0⃗, a⃗);
  3. return _mm_add_epi64 (b⃗, _mm_and_si128 (b⃗, p⃗)); }
```

If $m=n$, we can proceed as follows: $\max(x, p-y)$ equals $x$ if, and only if, $x \geqslant p-y$. In the latter case $(x+y) \operatorname{rem} p$ can be obtained as $x-(p-y)$, and otherwise as $x-(p-y)+p$. If $p=2^n$, an overflow only occurs when $y=0$. Nevertheless, `max (x, p - y)` equals $x$ when computed in U and `x - (p - y)` is the correct value. These calculations can be vectorized for packed integers of bit-size 8, 16, and 32 as follows:

**Function 4**

**Input.** Packed 32-bit integers $\vec{x}$ and $\vec{y}$ modulo $\vec{p}$.
**Output.** $(\vec{x}+\vec{y}) \operatorname{rem} \vec{p}$.

```
__m128i add_mod_epu32 (__m128i x⃗, __m128i y⃗) {
  1. __m128i a⃗ = _mm_sub_epi32 (p⃗, y⃗);
  2. __m128i b⃗ = _mm_cmpeq_epi32 (_mm_max_epu32 (x⃗, a⃗), x⃗);
  3. __m128i c⃗ = _mm_andnot_si128 (b⃗, p⃗);
  4. return _mm_add_epi32 (_mm_sub_epi32 (x⃗, a⃗), c⃗); }
```

The minimum operator and comparisons do not exist for packed 64-bit integers so we declare the function `_mm_cmpgt_epu64 (__m128i a⃗, __m128i b⃗)` as:

`_mm_cmpgt_epi64 (_mm_sub_epi64 (a⃗, `$\overrightarrow{2^{63}}$`), _mm_sub_epi64 (b⃗, `$\overrightarrow{2^{63}}$`))`

where $\overrightarrow{2^{63}}$ represents the packed 64-bit integer filled with $2^{63}$. The modular addition can be realized as follows:

**Function 5**

**Input.** Packed 64-bit integers $\vec{x}$ and $\vec{y}$ modulo $\vec{p}$.
**Output.** $(\vec{x}+\vec{y}) \operatorname{rem} \vec{p}$.

```
__m128i add_mod_epu64 (__m128i x⃗, __m128i y⃗) {
  1. __m128i a⃗ = _mm_add_epi64 (x⃗, y⃗);
  2. __m128i b⃗ = _mm_or_si128 (_mm_cmpgt_epu64 (x⃗, a⃗),
                               _mm_cmpgt_epu64 (a⃗, p⃗ - 1⃗));
  3. return _mm_sub_epi64 (a⃗, _mm_and_si128 (b⃗, p⃗)); }
```

It is straightforward to adapt these functions to the AVX 2 instruction set.

### 2.1.3. Timings

Table 1 displays timings for arithmetic operations over integer types of all possible bit-sizes $n$ supported by the compiler. Timings are the average number of clock cycles when applying the addition on two vectors of byte-size 4096 aligned in memory, and writing the result into the first vector. In particular, timings include load and store instructions. The row *Scalar* corresponds to disabling vectorization extensions with the command line option `-fno-tree-vectorize` of the compiler. For better performance, loops are unrolled and the optimization option `-O3` of the compiler is used. For conciseness, we restrict tables to addition; subtraction and negation behave in a similar way. Operations which are not supported are indicated by N/A, meaning *Non Applicable*. In the other rows of the table we indicate timings obtained by the vectorized algorithms according to the instruction set. Let us mention that the vectorization process is not left to the compiler. In fact we implemented a dedicated framework in Mathemagix, which is detailed in Section 4.

| $n$     | 8     | 16    | 32   | 64   | 128 |
|---------|-------|-------|------|------|-----|
| Scalar  | 1.6   | 1.5   | 1.7  | 1.6  | 4.0 |
| SSE 4.2 | 0.075 | 0.17  | 0.34 | 0.89 | N/A |
| AVX 2   | 0.044 | 0.093 | 0.18 | 0.59 | N/A |

**Table 1.** Sum of vectors of integers in CPU clock cycles.

Table 2 shows timings for modular sums. In absence of vectorization, 8-bit and 16-bit arithmetic operations are in fact performed by 32-bit operations. Indeed, 8-bit and 16-bit arithmetic is handled in a suboptimal way by current processors and compilers. For the vectorized implementations, we observe significant speedups when $m \leqslant n-1$. Nevertheless, when $m = n$, the lack of performance is not dramatic enough to justify the use of larger integers and double memory space.

| $n$     | 8     |      | 16   |      | 32   |      | 64  |     | 128 |     |
|---------|-------|------|------|------|------|------|-----|-----|-----|-----|
| $m$     | 7     | 8    | 15   | 16   | 31   | 32   | 63  | 64  | 127 | 128 |
| Scalar  | 2.3   | 2.3  | 2.4  | 2.4  | 2.4  | 2.5  | 2.6 | 2.9 | 12  | 16  |
| SSE 4.2 | 0.13  | 0.20 | 0.31 | 0.45 | 0.60 | 0.86 | 1.7 | 2.0 | N/A | N/A |
| AVX 2   | 0.081 | 0.12 | 0.16 | 0.23 | 0.31 | 0.44 | 1.1 | 1.6 | N/A | N/A |

**Table 2.** Modular sum in CPU clock cycles.

## 2.2. Barrett's product

The first modular product we describe is the one classically attributed to Barrett. It has the advantage to operate on integer types with no assumption on the modulus. For any nonnegative real number $x$, we write $\lfloor x \rfloor$ for the largest integer less or equal to $x$, and $\lceil x \rceil$ for the smallest integer greater or equal to $x$. We use the following auxiliary quantities:

- the nonnegative integer $r := \lceil \log p / \log 2 \rceil$ of $p$ with $2^{r-1} < p \leqslant 2^r$;
- nonnegative integers $s$ and $t$ such that $t \geqslant r$ and $s + t \leqslant n + r - 1$;
- the integer $q := \left\lfloor \frac{2^{s+t}}{p} \right\rfloor$ represents an approximation of a rescaled numerical inverse of $p$.

Since $2^{s+t}/p < 2^{n+r-1}/2^{r-1} = 2^n$, the integer $q$ fits in U. We call $q$ the *pre-inverse* of $p$. Since $s + t \leqslant 2n - 1$, the computation of $q$ just requires one division in L. In this subsection both integers $p$ and $q$ are assumed to be of type U. We describe below how to set suitable values for $s$ and $t$ in terms of $r$ (see Table 3).

### 2.2.1. Reduction

Let $\alpha$ be one more auxiliary quantity with $\alpha\, 2^t \leqslant 2^{2n}$. If $a$ is an integer such that $0 \leqslant a < \alpha\, p$, Barrett's algorithm computes $a \operatorname{rem} p$ as follows:

**Function 6**

    **Input.** An integer $a$ such that $0 \leqslant a < \alpha\, p$, with $p$, $q$, $r$, $s$, $t$ and $\alpha$ as above.
    **Output.** $a \operatorname{rem} p$.

```
U reduce_barrett (L a) {
 1. L b = a >> s;
 2. L c = (b * q) >> t;
 3. L d = a - c * p;
 4. while (d >= p) d = d - p;
 5. return d; }
```

PROPOSITION 1. *Function 6 is correct. The number of iterations in the "while" loop of step 4 is at most $h$, where $h := \left\lceil \frac{2^s}{2^{r-1}} + \frac{\alpha\, 2^r}{2^{s+t}} \right\rceil$ if $s \geqslant 1$ and $h := \left\lceil \frac{\alpha\, 2^r}{2^t} \right\rceil$ if $s = 0$. In addition, $b\, q$ and $c\, p$ fit in* L.

**Proof.** From $a < \alpha\, p$, it follows that $b\, q \leqslant a\, q / 2^s < \alpha\, p\, q / 2^s \leqslant \alpha\, 2^t$. Therefore, $b\, q$ has bit-size at most $2n$, and $c < \alpha$. Since $r \leqslant t$ the product $c\, p$ fits in L. From

$$c - \left\lfloor \frac{a}{p} \right\rfloor = \left( \left\lfloor \frac{b\, q}{2^t} \right\rfloor - \frac{b\, q}{2^t} \right) + \frac{q}{2^t} \left( b - \frac{a}{2^s} \right) + \frac{a}{2^{s+t}} \left( q - \frac{2^{s+t}}{p} \right) + \left( \frac{a}{p} - \left\lfloor \frac{a}{p} \right\rfloor \right),$$

we obtain $-1 - \frac{q}{2^t} - \frac{a}{2^{s+t}} < c - \left\lfloor \frac{a}{p} \right\rfloor < 1$. If $s \geqslant 1$, then $\frac{q}{2^t} + \frac{a}{2^{s+t}} \leqslant h$, so that $-h \leqslant c - \left\lfloor \frac{a}{p} \right\rfloor \leqslant 0$, whence the conclusion follows. If $s = 0$, then $b = a$, and we still have $-h \leqslant c - \left\lfloor \frac{a}{p} \right\rfloor \leqslant 0$. □

If $z$ is an integer modulo $p$, and $(x_i)_{i \in \{1,\ldots,l\}}$ and $(y_i)_{i \in \{1,\ldots,l\}}$ are sequences of integers modulo $p$, computing $(z + x_1\, y_1 + \cdots + x_l\, y_l) \bmod p$ is a central task to matrix and polynomial products. In order to minimize the number of reductions to be done, we wish to take $\alpha$ as large as possible. In general, in Barrett's algorithm, we can always take $\alpha = 2^n$, $t = n$ and $s = r - 1$, which leads to $h = 3$. When $r \leqslant n - 1$, we can achieve $h = 2$ if we restrict to $\alpha = 2^{n-1}$ with $t = n$ and $s = r - 1$. When $2 \leqslant r \leqslant n - 2$, it is even better to take $t = n + 1$ and $s = r - 2$ so that $h = 1$ when $\alpha = 2^{n-2}$. When $r \leqslant 1$, then we let $t = n + 1$ and $s = 0$ when $\alpha = 2^{n-2}$ to get $h = 1$. These possible settings are summarized in Table 3.

|   | $m \leqslant n - 2$ | $m \leqslant n - 1$ | $m \leqslant n$ |
|---|---|---|---|
| $\alpha$ | $2^{n-2}$ | $2^{n-1}$ | $2^n$ |
| $s$ | $\max\,(r-2,\,0)$ | $r-1$ | $r-1$ |
| $t$ | $n+1$ | $n$ | $n$ |
| $h$ | 1 | 2 | 3 |

**Table 3.** Values for $\alpha$, $s$, $t$, and $h$.

We could consider taking $\bar{q} := \lceil 2^{s+t} / p \rceil$ instead of $q$. From $s + t \leqslant n + r - 1$ and $2^n \geqslant 2^{r-1} + 1$ we obtain $2^{s+t} \leqslant 2^{n+r-1} \leqslant 2^{n+r-1} + 2^n - 2^{r-1} - 1 = (2^n - 1)(2^{r-1} + 1)$. Therefore the inequalities $\frac{2^{s+t}}{p} \leqslant \frac{2^{s+t}}{2^{r-1}+1} \leqslant 2^n - 1$ imply that $\bar{q}$ fits in U. Using $\bar{q}$ instead of $q$ in Function 6 leads to the following inequalities:

$$-1 - \frac{\bar{q}}{2^t} < c - \left\lfloor \frac{a}{p} \right\rfloor < \frac{a}{2^{s+t}} + 1 - \frac{1}{p}.$$

If $s=0$, then the term $\frac{\bar{q}}{2^t}$ disappears. If $r$ is sufficiently small, then $\frac{a}{2^{s+t}}$ can be bounded by $\frac{1}{p}$. Therefore line 4 of Function 6 can be discarded. More precisely, if we assume that $m \leqslant (n-1)/2$ and letting $s=0$, $t:=n+r-1$, and $\bar{q}:=\lceil 2^t/p \rceil$, then we have the following faster function:

**Function 7**

    **Input.** An integer $a$ with $0 \leqslant a < 2^{(n-1)/2} p$, and $p$, $\bar{q}$, $t = n+r-1$, $m \leqslant \frac{n-1}{2}$ as above.
    **Output.** $a \operatorname{rem} p$.

```
U reduce_barrett_half (U a) {
 1. U c = (a * ((L) q̄)) >> t;
 2. return a - c * p; }
```

PROPOSITION 2. *Function 7 is correct.*

**Proof.** Letting $\alpha := 2^{(n-1)/2}$, the proof follows from $-1 < c - \left\lfloor \frac{a}{p} \right\rfloor < \frac{a}{2^t} + 1 - \frac{1}{p}$ and $\frac{a p}{2^t} < \frac{\alpha p^2}{2^t} \leqslant \alpha \, 2^{2r-t} \leqslant \alpha \, 2^{r-n+1} \leqslant 1$. □

**Remark 3.** With recent compilers, 128-bit integers are available, which eases this implementation up to 64-bit. But if U is the largest type of integers supported by the compiler, then $q$ has to be computed by alternative methods such as the classical Newton-Raphson division [57] (see also our implementation in `numerix/modular_int.hpp` inside MATHEMAGIX, with the necessary proofs in the documentation).

### 2.2.2. Several products by the same multiplicand

Assume that we wish to compte several modular products, where one of the multiplicands is a fixed modular integer $y \in \mathbb{Z}/p\mathbb{Z}$. A typical application is the computation of FFTs, where we need to multiply by roots of unity that only depend on the size of the transformer and can thus be computed once and cached into memory. Then one can pre-compute $\psi := \lfloor (2^n y)/p \rfloor < 2^n$ and obtain a speed-up within each product by $y$ thanks to the following function:

**Function 8**

    **Input.** Integers $x$, $y$, and $\psi$ in $\{0, ..., p-1\}$, with $p$ and $\psi$ as above.
    **Output.** $(x y) \operatorname{rem} p$.

```
U mul_mod_barrett (U x, U y, U ψ) {
 1. U c = (x * ((L) ψ)) >> n;
 2. L d = ((L) x) * y - ((L) c) * p;
 3. return (d >= p) ? d - p : d; }
```

PROPOSITION 4. *Function 8 is correct.*

**Proof.** From $c - \left\lfloor \frac{x y}{p} \right\rfloor = \left( \left\lfloor \frac{x \psi}{2^n} \right\rfloor - \frac{x \psi}{2^n} \right) + \frac{x}{2^n} \left( \psi - \frac{2^n y}{p} \right) + \left( \frac{x y}{p} - \left\lfloor \frac{x y}{p} \right\rfloor \right)$ we obtain $-1 - \frac{x}{2^n} < c - \left\lfloor \frac{x y}{p} \right\rfloor < 1$, whence the correctness. □

Notice that Function 8 does not depend on $q$, $r$, $s$ or $t$. If $m \leqslant n-1$, then line 2 can be replaced by `U d = x * y - c * p`. If $m \leqslant n/2$, then $\bar{\psi} := \lceil (2^n y)/p \rceil$ fits in U since $(2^n y)/p \leqslant 2^n - 2^n/p$, and Function 8 can be improved along the same lines as Function 7:

**Function 9**

    **Input.** Integers $x$, $y$, $\bar{\psi}$ in $\{0, ..., p-1\}$, with $p$ and $\bar{\psi}$ as above, and where $m \leqslant n/2$.
    **Output.** $(x\,y) \operatorname{rem} p$.

```
U mul_mod_barrett_half (U x, U y, U ψ̄) {
  1. U c = (x * ((L) ψ̄)) >> n;
  2. return x * y - c * p; }
```

PROPOSITION 5. *Function 9 is correct.*

**Proof.** Similarly to previous proofs, we have $-1 < c - \left\lfloor \frac{x\,y}{p} \right\rfloor < \frac{x}{2^n} + 1 - \frac{1}{p}$, whence the correctness. □

### 2.2.3. Implementations with SSE 4.2 and AVX 2

Function 6 could be easily vectorized if all the necessary elementary operations were available within the SSE 4.2 or AVX 2 instruction sets. Unfortunately this is not so for all integer sizes, which forces us to examine different cases. As a first remark, in order to remove the branching involved in line 4 of Function 6, we replace it by d= min (d, d - p) as many times as specified by Proposition 1. In the functions below we consider the case when $m > n/2$. The other case is more straightforward. The packed $2\,n$-bit integer filled with $q$ (resp. $p$) seen of type L is written $\vec{q}_{\mathsf{L}}$ (resp. $\vec{p}_{\mathsf{L}}$). The constant $\overrightarrow{2^n - 1}_{\mathsf{L}}$ corresponds to the packed $2\,n$-bit integer filled with $2^n - 1$. We start with the simplest case of packed 16-bit integers. If $\vec{x}$ and $\vec{y}$ are the two vectors of scalar type U to be multiplied modulo $p$, then we unpack each of them into two vectors of scalar type L, we perform all the needed SIMD arithmetic over L, and then we pack the result back to vectors over U.

**Function 10**

    **Input.** Packed 16-bit integers $\vec{x}$ and $\vec{y}$ modulo $\vec{p}$, assuming $m \leqslant n - 2$.
    **Output.** $(\vec{x}\,\vec{y}) \operatorname{rem} \vec{p}$.

```
__m128i mul_mod_2_epu16 (__m128i x⃗, __m128i y⃗) {
  1.  __m128i x⃗_l = _mm_unpacklo_epi16 (x⃗, 0⃗);
  2.  __m128i x⃗_h = _mm_unpackhi_epi16 (x⃗, 0⃗);
  3.  __m128i y⃗_l = _mm_unpacklo_epi16 (y⃗, 0⃗);
  4.  __m128i y⃗_h = _mm_unpackhi_epi16 (y⃗, 0⃗);
  5.  __m128i a⃗_l = _mm_mullo_epi32 (x⃗_l, y⃗_l);
  6.  __m128i a⃗_h = _mm_mullo_epi32 (x⃗_h, y⃗_h);
  7.  __m128i b⃗_l = _mm_srli_epi32 (a⃗_l, s);
  8.  __m128i b⃗_h = _mm_srli_epi32 (a⃗_h, s);
  9.  __m128i c⃗_l = _mm_srli_epi32 (_mm_mullo_epi32 (b⃗_l, q⃗_L), t);
  10. __m128i c⃗_h = _mm_srli_epi32 (_mm_mullo_epi32 (b⃗_h, q⃗_L), t);
  11. __m128i c⃗   = _mm_packus_epi32 (c⃗_l, c⃗_h);
  12. __m128i d⃗   = _mm_sub_epi16 (_mm_mullo_epi16 (x⃗, y⃗),
                                   _mm_mullo_epi16 (c⃗, p⃗));
  13. return _mm_min_epu16 (d⃗, _mm_sub_epi16 (d⃗, p⃗));
```

If $m \leqslant n - 1$, then the computations up to line 10 are the same but the lines after are replaced by:

```
  11. __m128i d⃗_l = _mm_sub_epi32 (a⃗_l, _mm_mullo_epi32 (c⃗_l, p⃗_L));
  12. __m128i d⃗_h = _mm_sub_epi32 (a⃗_h, _mm_mullo_epi32 (c⃗_h, p⃗_L));
```

```
13. d⃗_l = _mm_min_epu32 (d⃗_l , _mm_sub_epi32 (d⃗_l , p⃗_L));
14. d⃗_h = _mm_min_epu32 (d⃗_h , _mm_sub_epi32 (d⃗_h , p⃗_L));
15. __m128i d⃗ = _mm_packus_epi32 (d⃗_l , d⃗_h);
16. return _mm_min_epu16 (d⃗, _mm_sub_epi16 (d⃗, p⃗));
```

Under the only assumption that $m \leqslant n$, ones needs to duplicate latter lines 13 and 14.

For packed 8-bit integers since no packed product is natively available, we simply perform most of the computations over packed 16-bit integers as follows:

**Function 11**

    **Input.** Packed 8-bit integers $\vec{x}$ and $\vec{y}$ modulo $\vec{p}$, assuming $m \leqslant n-2$.
    **Output.** $(\vec{x}\,\vec{y})\,\mathrm{rem}\,\vec{p}$.

```
__m128i muladd_mod_2_epu8 (__m128i x⃗, __m128i y⃗) {
 1. __m128i x⃗_l = _mm_unpacklo_epi8 (x⃗, 0⃗);
 2. __m128i x⃗_h = _mm_unpackhi_epi8 (x⃗, 0⃗);
 3. __m128i y⃗_l = _mm_unpacklo_epi8 (y⃗, 0⃗);
 4. __m128i y⃗_h = _mm_unpackhi_epi8 (y⃗, 0⃗);
 5. __m128i a⃗_l = _mm_mullo_epi16 (x⃗_l , y⃗_l );
 6. __m128i a⃗_h = _mm_mullo_epi16 (x⃗_h, y⃗_h);
 7. __m128i b⃗_l = _mm_srli_epi16 (a⃗_l , s);
 8. __m128i b⃗_h = _mm_srli_epi16 (a⃗_h, s);
 9. __m128i c⃗_l = _mm_srli_epi16 (_mm_mullo_epi16 (b⃗_l , q⃗_L), t);
10. __m128i c⃗_h = _mm_srli_epi16 (_mm_mullo_epi16 (b⃗_h, q⃗_L), t);
11. __m128i d⃗_l = _mm_sub_epi16 (a⃗_l , _mm_mullo_epi16 (c⃗_l , p⃗_L));
12. __m128i d⃗_h = _mm_sub_epi16 (a⃗_h, _mm_mullo_epi16 (c⃗_h, p⃗_L));
13. __m128i d⃗   = _mm_packus_epi16 (d⃗_l , d⃗_h);
14. return _mm_min_epu16 (d⃗, _mm_sub_epi8 (d⃗, p⃗));
```

If $m \leqslant n-2$ does not hold, then the same modifications as with packed 16-bit integers are applied.

For packed 32-bit integers, a similar extension using packing and unpacking instructions is not possible. We take advantage of the `_mm_mul_epu32` instruction. If $m \leqslant n-2$, then we use the following code:

**Function 12**

    **Input.** Packed 32-bit integers $\vec{x}$, $\vec{y}$ modulo $\vec{p}$, assuming $m \leqslant n-2$.
    **Output.** $(\vec{x}\,\vec{y})\,\mathrm{rem}\,\vec{p}$.

```
__m128i mul_mod_2_epu32 (__m128i x⃗, __m128i y⃗) {
 1. __m128i a⃗_l = _mm_mul_epu32 (x⃗, y⃗);
 2. __m128i b⃗_l = _mm_srli_epi64 (a⃗_l , s);
 3. __m128i c⃗_l = _mm_srli_epi64 (_mm_mullo_epi32 (b⃗_l , q⃗_L), t);
 4. __m128i x⃗_h = _mm_srli_si128 (x⃗, 4);
 5. __m128i y⃗_h = _mm_srli_si128 (y⃗, 4);
 6. __m128i a⃗_h = _mm_mul_epu32 (x⃗_h, y⃗_h);
 7. __m128i b⃗_h = _mm_srli_epi64 (a⃗_h, s);
 8. __m128i c⃗_h = _mm_srli_epi64 (_mm_mullo_epi32 (b⃗_h, q⃗_L), t);
 9. __m128i a⃗   = _mm_blend_epi16 (a⃗_l, _mm_slli_si128 (a⃗_h, 4),4+8+64+128);
10. __m128i c⃗   = _mm_or_si128 (c⃗_l , _mm_slli_si128 (c⃗_h, 4));
```

11. `__m128i` $\vec{d}$ = `_mm_sub_epi32` ($\vec{a}$, `_mm_mullo_epi32` ($\vec{c}$, $\vec{p}$));

12. `return` `_mm_min_epu32` ($\vec{d}$, `_mm_sub_epi32` ($\vec{d}$, $\vec{p}$));

When $m \leqslant n-1$, the same kind of modifications as before have to be done: $\vec{d}$ must be computed with vectors of unsigned 64-bit integers. Since these vectors do not support the computation of the minimum, one has to use `_mm_cmpgt_epi64`.

### 2.2.4. Timings

In Table 4 we display timings for multiplying machine integers, using the same conventions as in Table 1. Recall that packed 8-bit integers have no dedicated instruction for multiplication: it is thus done through the 16-bit multiplication *via* unpacking/packing.

| $n$ | 8 | 16 | 32 | 64 | 128 |
|---|---|---|---|---|---|
| Scalar | 1.8 | 1.6 | 1.6 | 1.6 | 5.3 |
| SSE 4.2 | 0.20 | 0.17 | 0.55 | N/A | N/A |
| AVX 2 | 0.11 | 0.10 | 0.27 | N/A | N/A |

**Table 4.** Product of integers in CPU clock cycles.

Table 5 shows the performance of the above algorithms. The row "Naive" corresponds to the scalar approach using the C++ operator `%` to compute remainders. Up to 32-bit integers, arithmetic is handled *via* 32 and 64-bit registers. The row "Barrett" contains timings for the scalar implementation of Barrett's product. Notice that the case $m = n - 2$ is significantly faster than the case $m = n$.

In the scalar approach, compiler optimizations and hardware operations are not always well supported for small sizes, so it makes sense to perform computations on a larger size. On our test platform, 32-bit integers typically provide the best performance. The resulting timings are given in the row "padded Barrett". For 8-bit integers, the best strategy is in fact to use lookup tables yielding 2.4 cycles in average, but this strategy cannot be vectorized. Finally the last two rows correspond to the vectorized versions of Barrett's approach.

| $n$ | 8 | | | | 16 | | | | 32 | | | |
|---|---|---|---|---|---|---|---|---|---|---|---|---|
| $m$ | 2 | 6 | 7 | 8 | 6 | 14 | 15 | 16 | 14 | 30 | 31 | 32 |
| Naive | 8.1 | 8.1 | 8.1 | 8.1 | 9.4 | 10 | 10 | 10 | 9.4 | 23 | 23 | 23 |
| Barrett | 3.4 | 8.5 | 9.7 | 13 | 3.2 | 6.2 | 8.3 | 10 | 3.2 | 6.1 | 8.3 | 10.6 |
| Padded Barrett | 3.1 | 3.1 | 3.1 | 3.1 | 3.1 | 3.1 | 3.1 | 5.9 | | | | |
| SSE Barrett | 0.78 | 0.78 | 0.86 | 1.1 | 0.76 | 1.9 | 2.4 | 2.7 | 2.3 | 3.3 | 3.6 | 7.1 |
| AVX Barrett | 0.40 | 0.40 | 0.45 | 0.54 | 0.41 | 1.0 | 1.2 | 1.4 | 1.3 | 1.8 | 2.1 | 4.0 |

**Table 5.** Modular product in CPU clock cycles.

For larger integers, the performance is shown in Table 6. Let us mention that in the row "Barrett" with $n = 64$ we actually make use of `__int128` integer arithmetic.

| $n$ | 64 | | | | 128 | | | |
|---|---|---|---|---|---|---|---|---|
| $m$ | 30 | 62 | 63 | 64 | 62 | 126 | 127 | 128 |
| Naive | 22 | 91 | 91 | 91 | 92 | 520 | 710 | 500 |
| Barrett | 8.2 | 15 | 25 | 32 | 71 | 98 | 144 | 154 |

**Table 6.** Modular product in CPU clock cycles.

Table 7 shows the average cost of one product with a fixed multiplicand. In comparison with Table 5, we notice a significant speedup, especially for the vectorial algorithms.

| $n$ | \multicolumn{2}{c}{8} | \multicolumn{2}{c}{16} | \multicolumn{2}{c}{32} | \multicolumn{2}{c}{64} | \multicolumn{2}{c}{128} |
|---|---|---|---|---|---|---|---|---|---|---|
| $m$ | 7 | 8 | 15 | 16 | 31 | 32 | 63 | 64 | 127 | 128 |
| Padded Barrett | 3.0 | 3.0 | 3.0 | 3.0 | 3.9 | 3.8 | 3.9 | 8.3 | 36 | 110 |
| SSE Barrett | 0.59 | 0.59 | 0.52 | 1.9 | 1.8 | 4.6 | | | | |
| AVX Barrett | 0.38 | 0.39 | 0.36 | 0.94 | 1.1 | 2.4 | | | | |

**Table 7.** Modular product for a fixed multiplicand in CPU clock cycles.

**Remark 6.** Our vectorized modular operations can easily be adapted to simultaneously compute modulo several moduli, say $p_1, p_2, p_3, p_4$, assuming they share the same parameters $r$, $s$ and $t$. Nevertheless, instead of using vectors $\vec{p}_\mathrm{L}$ filled with the same modulus, this requires one vector of type L to be filled with $p_1$ and $p_3$ and a second one with $p_2$ and $p_4$; the same consideration holds for $\vec{q}_\mathrm{L}$. These modifications involve to cache more precomputations and a small overhead in each operation.

## 2.3. Montgomery's product

For Montgomery's algorithm [52], one needs to assume that $p$ is odd. Let $m$ be such that $r \leqslant m \leqslant n$ and let $a < 2^m p$. We need the auxiliary quantities $\rho$ and $\chi$ defined by $0 < \chi < 2^m$, $0 < \rho < p$, and $\rho\, 2^m - \chi\, p = 1$. They can be classically computed with the extended Euclidean algorithm [26, Chapter 3].

### 2.3.1. Reduction

For convenience we introduce $\mu := 2^m - 1$. The core of Montgomery's algorithm is the following reduction function:

**Function 13**

    **Input.** An integer $a$ such that $0 \leqslant a < 2^m p$, with $p$ odd, and $m$ as above.
    **Output.** $(a\,\rho) \operatorname{rem} p$.

```
U reduce_montgomery (U a) {
1. U  b = (a * χ) & μ;
2. L  c = a + b * p;
3. U  d = c >> m;
4. if (c < a) return d - p;
5. return (d >= p) ? d - p : d; }
```

PROPOSITION 7. *Function 13 is correct. If $m \leqslant n-1$, then line 4 can be discarded.*

**Proof.** First one verifies that $b\, p \bmod 2^m = a\, \chi\, p \bmod 2^m = -a \bmod 2^m$. Therefore $a + b\, p$ is a multiple of $2^m$. If $m \leqslant n-1$, then no overflow occurs in the sum of line 2, and the division in line 3 is exact. We then have $d\, 2^m \bmod p = a \bmod p$, and the correctness follows from $0 \leqslant c < 2^m p + 2^m p$ and thus $0 \leqslant d < 2\, p$. If $m = n$, then the casual overflow in line 2 is tested in line 4, and $d$ is the value in U of $(a + b\, p)/2^m$. $\square$

Let $x$ be a modulo $p$ integer. We say that $x$ is in *Montgomery's representation* if stored as $(x\, 2^m) \operatorname{rem} p$. The product of two modular integers $x$ and $y$, of respective Montgomery's representations $\tilde{x}$ and $\tilde{y}$, can be obtained in Montgomery's representation $x\, y\, 2^m \operatorname{rem} p$ by applying Function 13 to $\tilde{x}\, \tilde{y}$ since $x\, y\, 2^m \bmod p = (x\, 2^m)\, (y\, 2^m)\, \rho \bmod p$.

If $m = n$, then the mask in line 1 can be avoided, and the shift in line 3 might be more favorable than a general shift, according to the compiler. In total, if $m = n$ or $m = n - 1$, Montgomery's approach is expected to be faster than Barrett's one. Otherwise costs should be rather similar. Of course these cost considerations are rather informal and the real cost very much depends on the processor and the compiler.

**Remark 8.** As for Barrett's algorithm one could be interested in simplifying Montgomery's product when performing several products by the same multiplicand $y$. Writing $\varphi = (\chi\, y) \operatorname{rem} 2^m$, the only simplification appears in line 1, where $b$ can be obtained as $(x\, \varphi) \operatorname{rem} 2^m$, which saves one product in U. Therefore Barrett's approach is expected to be always faster for this task. Precisely, if $m \leqslant n-1$, this is to be compared to Function 8 which performs only one high product in line 1.

**Remark 9.** Up to our best knowledge, it is not known whether Montgomery's product can be improved in a similar way as we did for Barrett's product, in the case when $m \leqslant n/2$.

### 2.3.2. Timings

Table 8 contains timings measured in the same manner as in the previous subsection. Compared to Tables 5 and 6 we observe that Montgomery's product is not interesting in the scalar case for 8 and 16-bit integers but it outperforms Barrett's approach in larger sizes, especially when $m = n$. Compared to Table 7, Montgomery's product is only faster for when $n = m = 64$ and $n = m = 128$.

| $n$ | 8 | | 16 | | 32 | | 64 | | 128 | |
|---|---|---|---|---|---|---|---|---|---|---|
| $m$ | 7 | 8 | 15 | 16 | 31 | 32 | 63 | 64 | 127 | 128 |
| Montgomery | 6.0 | 6.3 | 5.3 | 5.9 | 5.3 | 5.5 | 7.4 | 7.2 | 92 | 92 |
| SSE Montgomery | 0.75 | 1.1 | 2.2 | 3.2 | 3.4 | 5.0 | | | | |
| AVX Montgomery | 0.41 | 0.59 | 1.2 | 1.7 | 1.8 | 2.8 | | | | |

**Table 8.** Montgomery's product in CPU clock cycles.

## 3. Modular operations *via* numeric types

Instead of integer types, we can use numeric types such as `float` or `double` to perform modular operations. Let us write F such a type, and let $\ell \geqslant 3$ represent the size of the mantissa of F minus one, i.e. 23 bits for `float` and 52 bits for `double`. This number of bits corresponds to the size of the so called *trailing significant field* of F, which is explicitly stored. The modular product of $x \bmod p$ and $y \bmod p$ begins with the computation of an integer approximation $c$ of $x\,y/p$. Then $d = x\,y - c\,p$ is an approximation of $x\,y \operatorname{rem} p$ at distance $O(p)$. The constant hidden behind the latter $O$ depends on rounding modes. In this section we conform to IEEE-754 standard. We first analyze the case when $p$ fills less than half of the mantissa. We next propose a general algorithm using the fused multiply-add operation.

### 3.1. Notations

The following notations are used until the end of the present section. We write $u$ for the value of $1/p$ computed in F by applying the division operator on `1.0` and `(F) `$p$. Still following IEEE-754, the trailing significant field of $u$ is written $v$ and its exponent $e$. These quantities precisely depend on the active rounding mode used to compute $u$. But for all rounding modes, they satisfy the following properties:

$$0 \leqslant v < 2^\ell, \qquad u = \frac{2^\ell + v}{2^e}, \qquad \left|\frac{1}{p} - u\right| < \frac{1}{2^e}. \tag{1}$$

From the latter inequality we obtain

$$2^\ell - 1 \leqslant -1 + 2^\ell + v < \frac{2^e}{p} < 1 + 2^\ell + v \leqslant 2^{\ell+1},$$

and thus

$$\frac{p}{2^e} < \frac{1}{2^\ell - 1}. \tag{2}$$

Let $a$ be a real number of type `F`, and let `F b = a * u` be the approximation of $a\,u$ computed in `F`. Again, independently of the rounding mode, there exist integers $\beta$ and $f$, for the trailing significant field and the exponent of $b$, such that

$$0 \leqslant \beta < 2^\ell, \qquad b = \frac{2^\ell + \beta}{2^f}, \qquad |a\,u - b| < \frac{1}{2^f}. \tag{3}$$

From $2^\ell - 1 < 2^f a\,u \leqslant 2^{\ell+1}$ we deduce

$$\frac{1}{2^f} < \frac{a\,u}{2^\ell - 1}. \tag{4}$$

We also use the approximation $\bar{u}$ of `1.0 / ((F) p)` computed in `F` with rounding mode set towards infinity, so that $\bar{u} \geqslant 1/p$ holds.

### 3.2. Reduction in half size

When $p$ fills no more than half of the mantissa, that is when $m \leqslant \ell/2$, it is possible to perform modular products in `F` easily. Let the `floor` function return the largest integral value less than or equal to its argument, as defined in C99.

**Function 14**

   **Input.** An integer $a$ such that $0 \leqslant a < \alpha\,p$, where $\alpha := 2^{\ell/2}$, and $m \leqslant \ell/2$.
   **Output.** $a \operatorname{rem} p$.

```
F reduce_numeric_half (F a) {
  1. F b = a * u;
  2. F c = floor (b);
  3. F d = a - c * p;
  4. if (d >= p) return d - p;
  5. if (d < 0) return d + p;
  6. return d; }
```

PROPOSITION 10. *Function 14 is correct for any rounding mode. In addition, if $\bar{u}$ is used instead of $u$ in line 1, and if the rounding mode is set towards infinity, then line 4 can be discarded.*

**Proof.** Using (1) and (2) we obtain $a\,u = \frac{a}{p} u\,p < \alpha \left(1 + \frac{p}{2^e}\right) < 2^{\ell/2} \left(1 + \frac{1}{2^\ell - 1}\right)$, hence the `floor` function actually returns $\lfloor b \rfloor$ in $c$. From the decomposition

$$c - \left\lfloor \frac{a}{p} \right\rfloor = (\lfloor b \rfloor - b) + (b - a\,u) + a\left(u - \frac{1}{p}\right) + \left(\frac{a}{p} - \left\lfloor \frac{a}{p} \right\rfloor\right), \tag{5}$$

we deduce

$$-1 - \frac{1}{2^f} - \frac{\alpha\,p}{2^e} < c - \left\lfloor \frac{a}{p} \right\rfloor < \frac{1}{2^f} + \frac{\alpha\,p}{2^e} + 1.$$

From (2), we have $\frac{\alpha\,p}{2^e} < 1/2$, and from (4) we deduce $\frac{1}{2^f} < \frac{a\,u}{2^\ell - 1} \leqslant \frac{\alpha\,p\,(1/p + 1/2^e)}{2^\ell - 1} \leqslant \frac{\alpha + \alpha\,p/2^e}{2^\ell - 1} \leqslant \frac{2^{\ell/2} + 1/2}{2^\ell - 1} < \frac{1}{2}$. It follows that $\left|c - \left\lfloor \frac{a}{p} \right\rfloor\right| < 2$ hence that $\left|c - \left\lfloor \frac{a}{p} \right\rfloor\right| \leqslant 1$.

If using $\bar{u}$ instead of $u$, and if the rounding mode is set towards infinity, then $b \geqslant a\,\bar{u}$, and then $0 \leqslant c - \left\lfloor \frac{a}{p} \right\rfloor \leqslant 1$, which allows us to discard line 4. □

In the same way we did for Barrett's product, and under mild assumptions, we can improve the latter function whenever $m \leqslant (\ell - 1)/2$.

**Function 15**

    **Input.** An integer $a$ such that $0 \leqslant a < \alpha\, p$, where $\alpha := 2^{(\ell-1)/2}$ and $m \leqslant (\ell-1)/2$.
    **Hypothesis.** The current rounding mode rounds towards infinity.
    **Output.** $a \operatorname{rem} p$.

    F `reduce_numeric_half_1` (F $a$) {
      1. F $b$ = $a$ * $\bar{u}$;
      2. F $c$ = `floor` ($b$);
      3. `return` $a$ - $c$ * $p$; }

PROPOSITION 11. *Function 15 is correct.*

**Proof.** We claim that
$$\frac{1}{2^f} + \frac{\alpha\, p - 1}{2^e} \leqslant \frac{1}{p}. \tag{6}$$

By (2) and (4), the claim is satisfied as soon as $\frac{(\alpha\, p - 1)(1 + p/2^e)}{2^\ell - 1} + \frac{\alpha\, p - 1}{2^\ell - 1} \leqslant 1$, which is itself implied by $(2^{\ell-1} - 1)\left(1 + \frac{1}{2^\ell - 1}\right) + 2^{\ell-1} - 1 \leqslant 2^\ell - 1$, that is correct since it rewrites into $\frac{2^{\ell-1} - 1}{2^\ell - 1} \leqslant 1$ by expanding the product.

From $0 \leqslant \frac{a}{p} - \left\lfloor \frac{a}{p} \right\rfloor \leqslant 1 - \frac{1}{p}$ and (5) we deduce that
$$-1 < c - \left\lfloor \frac{a}{p} \right\rfloor < 1 + \frac{1}{2^f} + \frac{\alpha\, p - 1}{2^e} - \frac{1}{p} \leqslant 1,$$

which proves the correctness. □

**Remark 12.** The hypothesis on the rounding mode can be dropped if $a = 0$ or $p$ does not divide $a$. This is in particular case, if $p$ is prime and $a$ is the product of two numbers in $\{0, ..., p-1\}$. Indeed, if $a = 0$, then the algorithm is clearly correct. If $p$ does not divide $a$, then we have $\frac{1}{p} \leqslant \frac{a}{p} - \left\lfloor \frac{a}{p} \right\rfloor \leqslant 1 - \frac{1}{p}$. Combining the latter inequality with $-\frac{1}{2^f} < b - a\, u < \frac{1}{2^f}$ and $\left| a\, u - \frac{a}{p} \right| < \frac{\alpha\, p - 1}{2^f}$ yields
$$-1 - \frac{1}{2^f} - \frac{\alpha\, p - 1}{2^f} + \frac{1}{p} < b - \left\lfloor \frac{a}{p} \right\rfloor < 1 + \frac{1}{2^f} + \frac{\alpha\, p - 1}{2^e} - \frac{1}{p}.$$

Inequality (6) finally implies $-1 < b - \left\lfloor \frac{a}{p} \right\rfloor < 1$.

### 3.3. Larger modular products *via* FMA

Until now we have not exploited the whole mantissa of F. To release the assumption $m \leqslant \ell / 2$ in Function 14, the value for $d$ could be computed over a sufficiently large integer type. Over `double` one can use 64-bit integers, as implemented for instance in [58]. The drawbacks of this approach are the extra conversions between numeric and integer types, and the fact that the vectorization is compromised since 64-bit integer products are not natively available in the SSE or AVX instruction sets. In what follows we describe a modular product in the case when $m \leqslant \ell - 2$, using the fused multiply-add operation from the IEEE 754-2008 standard. We write `fma (x, y, z)` for the evaluation of $x\, y + z$ using the current rounding mode.

**Function 16**

**Input.** Integers $a_1$ and $a_2$ such that $0 \leqslant a_1 a_2 < \alpha p$, where $\alpha := 2^{\ell-2}$, and $m \leqslant \ell - 2$.
**Output.** $(a_1 a_2) \operatorname{rem} p$.

```
F mul_mod_fma (F a₁, F a₂) {
 1. F h = a₁ * a₂;
 2. F l = fma (a₁, a₂, −h);
 3. F b = h * u;
 4. F c = floor (b);
 5. F d = fma (−c, p, h);
 6. F e = d + l;
 7. if (e >= p) return e - p;
 8. if (e < 0) return e + p;
 9. return e; }
```

PROPOSITION 13. *Function 16 is correct for any rounding mode. If $\bar{u}$ is used instead of $u$ in line 3, and if the rounding mode is set towards infinity, then line 7 can be discarded.*

**Proof.** Let $a := a_1 a_2$. We have $|h - a| < 2^{2r-\ell} \leqslant 2^{\ell-4}$, so that $h + l = a$. We also verify that $a u = \frac{a}{p} u p < \alpha \left(1 + \frac{p}{2^e}\right) < 2^{\ell-2} \left(1 + \frac{1}{2^\ell - 1}\right)$, which implies that $c = \lfloor b \rfloor$. Using (5) again, the following inequalities hold:

$$-1 - \frac{1}{2^f} - \frac{\alpha p}{2^e} < c - \left\lfloor \frac{a}{p} \right\rfloor < \frac{1}{2^f} + \frac{\alpha p}{2^e} + 1.$$

From (2) we have $\frac{\alpha p}{2^e} \leqslant \frac{2^{\ell-2}}{2^\ell - 1} < \frac{1}{2}$, and from (4) we deduce $\frac{1}{2^f} \leqslant \frac{a u}{2^\ell - 1} \leqslant \frac{\alpha (1 + p/2^e)}{2^\ell - 1} < \frac{2^{\ell-2} + 1/2}{2^\ell - 1} \leqslant \frac{1}{2}$. It follows that $-1 \leqslant c \leqslant 1$. In particular this implies $|a_1 a_2 - c p| < 2 p$, whence $|h - c p| \leqslant l + 2 p < 2^\ell$. This proves that $d = h - c p$ and therefore $e = d + l = h - c p + l = a - c p$, which finally implies the correctness of Function 16.

Now suppose that $\bar{u}$ is used and that the rounding mode is set towards infinity. Then $\bar{u} \geqslant 1/p$, $h \geqslant a$ and $b \geqslant h \bar{u}$ so that $b \geqslant a \bar{u}$, and therefore $0 \leqslant c \leqslant 1$. □

**Remark 14.** Let `I` be a type of signed integers of at least $n > \ell$ bits. In the above functions the executable code generated for the `floor` instruction heavily depends on the compiler, its version, and its command line arguments. This makes timings for this numeric approach rather unpredictable. We have run tests with GCC version $\geqslant 4.7$ and CLANG [17] version $\geqslant 3.4$ and observed that `c = (F) ((I) b)` always generates `cvttsd2si`, `cvtsi2sd` instructions which are the x86 cast instructions, whereas `c = floor (b)` is compiled into a call to the `floor` function from the math library. In order to force the compiler to use a special purpose instruction such as `roundsd` from SSE 4.1, the `-O3 -msse4.1 -fno-trapping-math` arguments must be passed to `gcc`. In the case of `clang`, the options `-O3 -msse4.1` are sufficient.

### 3.4. Vectorized implementations

For efficiency, we assume that SSE 4.1 is available. Our modular addition and subtraction functions can benefit from `_mm_blendv_pd` as follows:

**Function 17**

> **Input.** Packed doubles $\vec{x}$ and $\vec{y}$ modulo $\vec{p}$, assuming $m \leqslant \ell - 1$.
> **Output.** $(\vec{x} + \vec{y})\,\text{rem}\,\vec{p}$.
>
> ```
> __m128d add_mod_1 (__m128d x⃗, __m128d y⃗) {
>   1. __m128d a⃗ = _mm_add_pd (x⃗, y⃗);
>   2. __m128d b⃗ = _mm_sub_pd (a⃗, p⃗);
>   3. return _mm_blendv_pd (b⃗, a⃗, b⃗); }
> ```

Assuming that FMA instructions are available, Function 16 is implemented as follows:

**Function 18**

> **Input.** Packed doubles $\vec{x}$ and $\vec{y}$ modulo $\vec{p}$, assuming $m \leqslant \ell - 2$.
> **Output.** $(\vec{x}\,\vec{y})\,\text{rem}\,\vec{p}$.
>
> ```
> __m128d mul_mod_1 (__m128d x⃗, __m128d y⃗) {
>   1. __m128d h⃗ = _mm_mul_pd (x⃗, y⃗);
>   2. __m128d l⃗ = _mm_fmsub_pd (x⃗, y⃗, h⃗);
>   3. __m128d b⃗ = _mm_mul_pd (h⃗, u⃗);
>   4. __m128d c⃗ = _mm_floor (b⃗);
>   5. __m128d d⃗ = _mm_fnmadd_pd (c⃗, p⃗, h⃗);
>   6. __m128d e⃗ = _mm_add_pd (d⃗, l⃗);
>   7. __m128d t⃗ = _mm_sub_pd (e⃗, p⃗);
>   8. e⃗ = _mm_blendv_pd (t⃗, e⃗, t⃗);
>   9. t⃗ = _mm_add_pd (e⃗, p⃗);
>  10. return _mm_blendv_pd (e⃗, t⃗, e⃗); }
> ```

Our AVX based implementation is the same, *mutatis mutandis*.

## 3.5. Timings

Table 9 is obtained under similar conditions as Tables 1 and 4, but for the types `float` and `double`. The row "Scalar" corresponds to the unvectorized implementation with unrolled loops, and preventing the compiler from using auto-vectorization. The next rows concern timings using SSE 4.1 and AVX instructions with unrolled loops on aligned data.

| Operation | sum | | product | |
|---|---|---|---|---|
| Type | `float` | `double` | `float` | `double` |
| Scalar | 1.1 | 1.1 | 1.1 | 1.1 |
| SSE | 0.32 | 1.0 | 0.30 | 1.0 |
| AVX | 0.19 | 0.55 | 0.16 | 0.55 |

**Table 9.** Floating point operations in CPU clock cycles.

In Table 10 we have shown timings for the modular sum and product functions from this section. The row "Scalar" excludes auto-vectorization and does not use the processor built-in FMA unit.

| Operation | sum | | product | | | | |
|---|---|---|---|---|---|---|---|
| Type | `float` | `double` | `float` | | `double` | | |
| $m$ | 21 | 50 | 11 | 21 | 25 | 26 | 50 |
| Scalar | 2.0 | 2.1 | 4.7 | 8.1 | 4.7 | 6.4 | 7.5 |
| SSE & FMA | 0.60 | 1.3 | 1.0 | 1.5 | 2.0 | 2.8 | 3.0 |
| AVX & FMA | 0.33 | 0.75 | 0.54 | 0.8 | 1.2 | 1.8 | 1.9 |

**Table 10.** Modular operations in CPU clock cycles.

We notice that the timings in Table 10 are interesting when compared to those in Table 5. However, for multiplications with a fixed multiplicand, the approach from Table 7 becomes more attractive, and this will indeed be used in Section 5 below.

We also notice that efficient hardware quadruple precision arithmetic would allow us to consider moduli with larger bit sizes. An alternative to hardware quadruple precision arithmetic would be to provide an efficient "three-sum" $a + b + c$ instruction with correct IEEE-754 rounding. This would actually allow for the efficient implementation of more general medium precision floating point arithmetic.

## 4. IMPLEMENTATION DESIGN IN C++

The C++ libraries of MATHEMAGIX are not only oriented towards mathematical interfaces but also towards reusability of generic low level implementations. Advanced users can not only build on existing algorithms, but also replace them *a posteriori* by more efficient implementations for certain special cases. We heavily rely on C++ template programming. This section surveys some of the design principles behind our implementation of modular numbers but also of polynomials, matrices and other objects in MATHEMAGIX. Our techniques turn out to be especially useful for implementing the scalar and SIMD algorithms in a unified framework.

### 4.1. Data structures and top level functions

Consider a typical template class in MATHEMAGIX, such as `modulus`, `vector`, `matrix`, `polynomial` or `series`. Besides the usual template parameters, such as a coefficient type `C`, such classes generally come with a special additional parameter `V`, called the *implementation variant*. The parameter `V` plays a similar role as *traits* classes in usual C++ terminology [1, Chapter 2]. The variant `V` does *not* impact the internal representation of instances of the class, but it does control the way in which basic functions manipulate such instances.

For example, the class `vector<C,V>` (defined in `basix/vector.hpp`) corresponds to dense vectors with entries in `C`, stored in a contiguous segment of memory. A vector of this type consists of a pointer of type `C*` to the first element of the vector, the size $n$ of the vector, and the size $l$ of the allocated memory. For the sake of simplicity we omit that our vectors are endowed with reference counters. At the top level user interface, for instance, the sum of two vectors is defined as follows:

```
template<typename C, typename V>
vector<C,V> operator + (const vector<C,V>& v, const vector<C,V>& w) {
  typedef implementation<vector_linear, V> Vec;
  nat n= N(v); nat l= aligned_size<C,V> (n);
  C* t= mmx_new<C> (l);
```

```
      Vec::add (t, seg (v), seg (w), n);
      return vector<C,V> (t, n, l); }
```
In this piece of code N(v) represents the size of v, aligned_size<C,V> (n) computes the length to be allocated in order to store vectors of size n over C in memory. According to the values of C and V, we can force the allocated memory segment to start at an address multiple of a given value, such as 16 bytes when using SSE instructions, or 32 bytes when AVX support is enabled. The function mmx_new<C> is a reimplementation in MATHEMAGIX of the usual new function, which allows faster allocation of objects of small sizes. Finally, the data t, n, l are stored into an instance of vector<C,V>. The allocated memory is released once the vector is not used anymore, i.e. when its reference counter becomes zero. The core of the computations is implemented in the static member function add of the class implementation<vector_linear,V> explained in the next paragraphs. At the implementation level, operations are usually performed efficiently on the object representations. The expression seg (v) returns the first address of type C* of the memory segment occupied by v.

## 4.2. Algorithms and implementations

The classes containing the implementations are specializations of the following class:
```
    template<typename F, typename V, typename W=V> struct implementation;
```
The first template argument F is usually an empty class which names a set of *functionalities*. In the latter example we introduced vector_linear, which stands for the usual entry-wise vector operations, including addition, subtraction, component-wise product, etc. The value of the argument V for *naive implementations* of vector operations is vector_naive. The role of the third argument W will be explained later. The naive implementation of the addition of two vectors is then declared as a static member of implementation<vector_linear, vector_naive> as follows:
```
    template<typename V>
    struct implementation<vector_linear, V, vector_naive> {
      static inline void
      add (C* dest, const C* s1, const C* s2, nat n) {
        for (nat i= 0; i < n; i++)
          dest[i]= s1[i] + s2[i]; }                                    ../..
```
Four by four *loop unrolling* can for instance be implemented within another variant, say vector_unrolled_4, as follows:
```
    template<typename V>
    struct implementation<vector_linear, V, vector_unrolled_4> {
      static inline void
      add (C* dest, const C* s1, const C* s2, nat n) {
        nat i= 0;
        for (; i + 4 < n; i += 4) {
          dest[i  ]= s1[i  ] + s2[i  ]; dest[i+1]= s1[i+1] + s2[i+1];
          dest[i+2]= s1[i+2] + s2[i+2]; dest[i+3]= s1[i+3] + s2[i+3]; }
        for (; i < n; i++)
          dest[i]= s1[i] + s2[i]; }                                    ../..
```
When defining a new variant we hardly ever want to redefine the whole set of functionalities of other variants. Instead we wish to introduce new algorithms for a subset of functionalities, and to have the remaining implementations inherit from other variants. We say that a variant V *inherits* from W when the following partial specialization is active:

```
template<typename F, typename U>
struct implementation<F,U,V>: implementation<F,U,W> {};
```
For instance, if the variant `V` inherits from `vector_naive`, then the `add` function from `implementation<vector_linear,V>` is inherited from `implementation<vector_linear, vector_naive>`, unless the following partial template specialization is implemented: `template<typename U> implementation<vector_linear,U,V>`.

It remains to explain the role of the three parameters of `implementation`. In fact the second parameter keeps track of the top level variant from which `V` inherits. Therefore, in a static member of `implementation<F,U,V>`, when one needs to call a function related to another set of functionalities `G`, then it is fetched in `implementation<G,U>`. In order to illustrate the interest of this method, let us consider polynomials in the next paragraphs.

Our class `polynomial<C,V>` (defined in `algebramix/polynomial.hpp`) represents polynomials with coefficients in `C` using implementation variant `V`. Each instance of a polynomial is represented by a vector, that is a pointer with a reference counter to a structure containing a pointer to the array of coefficients of type `C*` with its allocated size, and an integer for the length of the considered polynomial (defined as the degree plus one). The set of functionalities includes linear operations, mainly inherited from those of the vectors (since the internal representations are the same), multiplication, division, composition, Euclidean sequences, subresultants, evaluations, Chinese remaindering, etc. All these operations are implemented for the variant `polynomial_naive` (in the file `algebramix/polynomial_naive.hpp`) with the most general but naive algorithms (with quadratic cost for multiplication and division).

The variant `polynomial_dicho<W>` inherits from the parameter variant `W` and contains implementations of classical divide and conquer algorithms: Karatsuba for the product, Sieveking's polynomial division, half-gcd, divide and conquer evaluation and interpolations [26, Chapters 8–11]. Polynomial product functions belong to the set of functionalities `polynomial_multiply`, and division functions to the set `polynomial_divide`. The division functions of `polynomial_dicho<W>` are implemented as static members of
```
template<typename U, typename W>
struct implementation<polynomial_divide,U,polynomial_dicho<W> >
```
They make use of the product functions of `implementation<polynomial_multiply,U>`.

Let us now consider polynomials with modular integer coefficients and let us describe how the Kronecker substitution product is plugged in. In a nutshell, the coefficients of the polynomials are first lifted into integers. The lifted integer polynomials are next evaluated at a sufficiently large power of two, and we compute the product of the resulting integers. The polynomial product is retrieved by splitting the integer product into chunks and reducing them by the modulus. For details we refer the reader for instance to [26, Chapter 8]. As to our implementation, we first create the new variant `polynomial_kronecker<W>` on top of another variant `W` (see file `algebramix/polynomial_kronecker.hpp`), which inherits from `W`, but which only redefines the implementation of the product in
```
template<typename U, typename W>
struct implementation<polynomial_multiply,U,polynomial_kronecker<W> >
```
When using the variant `K` defined by
```
typedef polynomial_dicho<polynomial_kronecker<polynomiam_naive> > > K;
```
the product functions in `implementation<polynomial_multiply, K>` correspond to the Kronecker substitution. The functions in `implementation<polynomial_division, K>` are inherited from
```
implementation<polynomial_division, K,
               polynomial_dicho<polynomial_naive> >
```

and thus use Sieveking's division algorithm. The divisions rely in their turn on the multiplication from `implementation<polynomial_multiply, K>`, which benefits from Kronecker substitution. In this way, it is very easy for a user to redefine a new variant and override given functions *a posteriori*, without modifying existing code.

Finally, for a given mathematical template type, we define a *default variant* as a function of the remaining template parameters. For instance, the default variant of the parameter `V` in `vector<C,V>` is `typename vector_variant<C>::type` which is intended to provide users with reasonable performance. This default value is set to `vector_naive`, but can be overwritten for special coefficients. The default variant is also the default value of the variant parameter in the declaration of `vector`. Users can thus simply write `vector<C>`. The same mechanism applies to polynomials, series, matrices, etc.

### 4.3. Modular integers

In MATHEMAGIX, moduli are stored in the dedicated class `modulus<C,V>` (from `numerix/modulus.hpp`). This abstraction allows us to attach extra information to the modulus, such as a pre-inverse. Modular numbers are instances of `modular<M,W>` (from `numerix/modular.hpp`), where `M` is a modulus type, and `W` is a variant specifying the way how the modulus is stored (e.g. in a global variable, or as an additional field for each modular number). Scalar arithmetic over integer (resp. numeric) types is implemented in `numerix/modular_int.hpp` (resp. `numerix/modular_double.hpp`). For convenience, packed integer and numeric types are wrapped into C++ classes so that modular operations can easily be implemented for them. Let us mention that vectors of modular numbers have a specific implementation variant allowing to share the same modulus when performing entry-wise operations. In this way we avoid fetching the modulus for each arithmetic operation.

Operations on vectors of integer and numeric types are implemented in a hierarchy of variants. One major variant controls the way loops are unrolled. Another important variant is dedicated to memory alignement.

## 5. FAST FOURIER TRANSFORM AND APPLICATIONS

In order to benefit from vectorized modular arithmetic within the fast Fourier transform, we implemented a vectorized variant of the classical in-place algorithm. In this section, we describe this variant and its applications to polynomial and matrix multiplication.

### 5.1. Vectorized truncated Fourier transform

Let $K$ be a commutative field, let $n = 2^k$ with $k \in \mathbb{N}$, and let $\omega \in K$ be a primitive $n$-th root of unity, which means that $\omega^n = 1$, and $\omega^j \neq 1$ for all $j \in \{1, ..., n-1\}$. Let $\mathcal{A}$ be a $K$-vector space. The fast Fourier transform (with respect to $\omega$) of an $n$-tuple $(a_0, ..., a_{n-1}) \in \mathcal{A}^n$ is the $n$-tuple $(\hat{a}_0, ..., \hat{a}_{n-1}) =: \mathrm{FFT}_\omega(a) \in \mathcal{A}^n$ with

$$\hat{a}_i = \sum_{j=0}^{n-1} \omega^{ij} a_j.$$

In other words, $\hat{a}_i = A(\omega^i)$, where $A \in \mathcal{A} \otimes K[X]$ denotes the element $A(X) := \sum_{i=0}^{n-1} a_i \otimes X^i$. If $i \in \{0, ..., n-1\}$ has binary expansion $i_0 + i_1 2 + i_2 2^2 + \cdots + i_{k-1} 2^{k-1}$, then we write $[i]_k = i_{k-1} + i_{k-2} 2 + i_{k-3} 2^2 + \cdots + i_0 2^{k-1}$ for the *bitwise mirror* of $i$ in length $k$. Following the terminology of [40], the *truncated Fourrier transform* (TFT) of an $l$-tuple $(a_0, ..., a_{l-1}) \in \mathcal{A}^l$ (with respect to $\omega$) is

$$\mathrm{TFT}_\omega(a) := (\hat{a}_{[0]_k}, ..., \hat{a}_{[l-1]_k}) = (A(\omega^{[0]_k}), ..., A(\omega^{[l-1]_k})).$$

The classical FFT and TFT algorithms (see [37, 40], for instance) do not directly exploit low level vectorization. Let $n_1 = 2^{k_1} < l$ be a divisor of $n$. The following algorithm reduces one TFT of size $l$ over $K$ into one TFT over $K^{n_1}$ of size $\lambda := \lceil l/n_1 \rceil$ and into one TFT of size $n_1$ over $K^\lambda$.

**Algorithm 1**

    **Input.** $n = 2^k$, $\omega$ a $n$-th primitive root of unity, $l \leqslant n$, $(a_0, ..., a_{l-1}) \in K^l$, and an integer $n_1 = 2^{k_1} \leqslant l$. Let $n_2 = 2^{k_2} = n/n_1$.

    **Output.** $A\bigl(\omega^{[i]_k}\bigr)$ for all $i \in \{0, ..., (\lambda - 1) n_1 - 1\}$, where $\lambda := \lceil l/n_1 \rceil$, and $A(X) = a_0 + a_1 X + \cdots + a_{l-1} X^{l-1}$.

*For convenience we consider that $a_i = 0$ for all $i \in \{l, ..., n\}$.*

1. Let $b_{j_2} := (a_{j_2 n_1}, ..., a_{(j_2+1)n_1-1})$ for all $j_2 \in \{0, ..., \lambda-1\}$.
   Compute $\bigl(\hat{b}_{[0]_{k_2}}, ..., \hat{b}_{[\lambda-1]_{k_2}}\bigr) := \mathrm{TFT}_{\omega^{n_1}}(b_0, ..., b_{\lambda-1})$.

2. Let $c_{j_1} := \Bigl(\hat{b}_{[0]_{k_2},j_1} \omega^{j_1[0]_{k_2}}, \hat{b}_{[1]_{k_2},j_1} \omega^{j_1[1]_{k_2}}, ..., \hat{b}_{[\lambda-1]_{k_2},j_1} \omega^{j_1[\lambda-1]_{k_2}}\Bigr)$ for all $j_1 \in \{0, ..., n_1-1\}$.

3. Compute $\bigl(\hat{c}_{[0]_{k_1}}, ..., \hat{c}_{[n_1-1]_{k_1}}\bigr) := \mathrm{TFT}_{\omega^{n_2}}(c_0, ..., c_{n_1-1})$.

4. Return $\bigl(\hat{c}_{[0]_{k_1},0}, ..., \hat{c}_{[n_1-1]_{k_1},0}, \hat{c}_{[0]_{k_1},1}, ..., \hat{c}_{[n_1-1]_{k_1},1}, ..., \hat{c}_{[0]_{k_1},\lambda-1}, ..., \hat{c}_{[n_1-1]_{k_1},\lambda-1}\bigr)$.

**PROPOSITION 15.** *Algorithm 1 is correct and takes at most $\frac{3}{2} \lambda n_1 k + O(n)$ operations in $K$, assuming given all the powers of $\omega$.*

**Proof.** Let $B_i(X) = b_{0,i} + b_{1,i} X + \cdots + b_{\lambda-1,i} X^{\lambda-1}$ for $i \in \{0, ..., n_1-1\}$, so that we have $A(X) = B_0(X^{n_1}) + B_1(X^{n_1}) X + \cdots + B_{n_1-1}(X^{n_1}) X^{n_1-1}$ and $\hat{b}_{[j_2]_{k_2}} = \Bigl(B_0\bigl(\omega^{n_1[j_2]_{k_2}}\bigr), ..., B_{n_1-1}\bigl(\omega^{n_1[j_2]_{k_2}}\bigr)\Bigr)$ for all $j_2 \in \{0, ..., \lambda-1\}$. Let $j_1 \in \{0, ..., n_1-1\}$ and $j_2 \in \{0, ..., \lambda-1\}$. A straightforward calculation leads to $A\bigl(\omega^{[j_2 n_1 + j_1]_k}\bigr) = A\Bigl(\omega^{[j_1]_{k_1} n_2 + [j_2]_{k_2}}\Bigr) =$

$$
\begin{aligned}
& \hat{b}_{[j_2]_{k_2},0} + \hat{b}_{[j_2]_{k_2},1} \omega^{[j_1]_{k_1} n_2 + [j_2]_{k_2}} + \cdots + \hat{b}_{[j_2]_{k_2},n_1-1} \omega^{([j_1]_{k_1} n_2 + [j_2]_{k_2})(n_1-1)} \\
={}& \hat{b}_{[j_2]_{k_2},0} + \Bigl(\hat{b}_{[j_2]_{k_2},1} \omega^{[j_2]_{k_2}}\Bigr) \omega^{[j_1]_{k_1} n_2} + \cdots + \Bigl(\hat{b}_{[j_2]_{k_2},n_1-1} \omega^{[j_2]_{k_2}(n_1-1)}\Bigr) \omega^{[j_1]_{k_1} n_2 (n_1-1)} \\
={}& c_{0,j_2} + c_{1,j_2} \omega^{[j_1]_{k_1} n_2} + \cdots + c_{n_1-1,j_2} \omega^{[j_1]_{k_1} n_2 (n_1-1)} = \hat{c}_{[j_1]_{k_1},j_2}.
\end{aligned}
$$

By [40, Theorem 1] step 1 can be done with $\frac{3}{2} n_1 (\lambda k_2 + n_2) + n_1/2$ operations in $K$. Step 2 involves $\lambda n_1$ operations and step 3 takes $\frac{3}{2} \lambda (n_1 k_1 + n_1) + n_2/2$ more operations. $\square$

Inverting Algorithm 1 is straightforward: it suffices to invert steps from 4 to 1 and use the inverse of the TFT. If $l = n$, then Algorithm 1 can be used to compute the FFT. If $n_1$ is taken to be the size corresponding to a machine vector, then most of the TFT can be performed using SIMD operations. For sizes of order at most a few kilobytes we actually use this strategy. In larger sizes it is preferable to take $n_1$ much larger, for instance of order $\sqrt{n}$, so that the TFT runs on rather large vectors. With $n_1$ of order $\sqrt{n}$, this algorithm is very close to the cache-friendly version of the TFT designed in [36].

A critical point in large sizes becomes the matrix transposition, necessary to reorganize data in steps 1 and 4 of Algorithm 1. We designed *ad hoc* cache-friendly SIMD versions for it. Table 11 reports on timings obtained in this way for $K = \mathbb{Z}/p\mathbb{Z}$ with $p = 469762049 = 7 \times 2^{26} + 1$ in MATHEMAGIX, with Barrett's approach. All the necessary primitives roots and powers are pre-computed and cached in memory. We observe significant speed-ups for the SSE and AVX versions.

| $n$ | $2^8$ | $2^9$ | $2^{10}$ | $2^{11}$ | $2^{12}$ | $2^{13}$ | $2^{14}$ | $2^{15}$ | $2^{16}$ | $2^{17}$ | $2^{18}$ | $2^{19}$ | $2^{20}$ |
|---|---|---|---|---|---|---|---|---|---|---|---|---|---|
| Scalar | 3.0 | 6.7 | 15 | 33 | 73 | 160 | 340 | 730 | 1600 | 3300 | 7100 | 15000 | 32000 |
| SSE | 1.0 | 2.2 | 4.8 | 11 | 22 | 48 | 110 | 220 | 470 | 1100 | 2200 | 4700 | 11000 |
| AVX | 0.64 | 1.4 | 2.9 | 6.4 | 14 | 30 | 64 | 140 | 290 | 630 | 1400 | 3000 | 6800 |

**Table 11.** FFT of size $n$ over $\mathbb{Z}/469762049\,\mathbb{Z}$, user time in microseconds.

Table 12 concerns FFTs for $K = \mathbb{Z}/p\,\mathbb{Z}$ and $p = 1108307720798209$. We compare NTL 6.1.0 with our FFT implementation, relying on AVX and FMA.

| $n$ | $2^8$ | $2^9$ | $2^{10}$ | $2^{11}$ | $2^{12}$ | $2^{13}$ | $2^{14}$ | $2^{15}$ | $2^{16}$ | $2^{17}$ | $2^{18}$ | $2^{19}$ | $2^{20}$ |
|---|---|---|---|---|---|---|---|---|---|---|---|---|---|
| NTL | 2.2 | 4.5 | 9.8 | 21 | 51 | 110 | 250 | 540 | 1200 | 2400 | 5300 | 12000 | 30000 |
| Mathemagix | 1.2 | 2.5 | 5.6 | 12 | 26 | 57 | 120 | 270 | 580 | 1200 | 2600 | 6100 | 15000 |

**Table 12.** FFT of size $n$ over $\mathbb{Z}/1108307720798209\,\mathbb{Z}$, user time in microseconds.

For the sake of comparison, we also report on the performance of the FFT over complex numbers in double precision. Table 13 compares timings provided by the command `test/bench` bundled with FFTW version 3.3.4 [25] (configured with the `--enable-avx` option), to the Mathemagix implementation in `algebramix/fft_split_radix.hpp`.

| $n$ | $2^8$ | $2^9$ | $2^{10}$ | $2^{11}$ | $2^{12}$ | $2^{13}$ | $2^{14}$ | $2^{15}$ | $2^{16}$ | $2^{17}$ | $2^{18}$ | $2^{19}$ | $2^{20}$ |
|---|---|---|---|---|---|---|---|---|---|---|---|---|---|
| FFTW3 | 0.42 | 0.96 | 2.4 | 5.6 | 14 | 35 | 84 | 190 | 410 | 880 | 2500 | 8000 | 19000 |
| Mathemagix | 0.48 | 1.0 | 2.3 | 5.7 | 14 | 33 | 77 | 200 | 410 | 900 | 2100 | 6100 | 15000 |

**Table 13.** FFT for complex of `double` of size $n$, user time in microseconds.

## 5.2. Polynomial matrix product

One major application of the TFT is the computation of polynomial products. In Table 14 we provide timings for multiplying two polynomials of degrees $<d$ over $\mathbb{Z}/469762049\,\mathbb{Z}$. For this task, we compare NTL and FLINT 2.4.3 [34] to our implementations. The row "Kronecker" corresponds to the classical Kronecker substitution algorithm [26, Section 8.4] that reduces the product to one integer product, for which we appeal to GMP version 6.0.0a [31]. The last row concerns the AVX 2 version of our modular TFT. Here an important fact must be noticed. Previous timings of TFT were obtained as average on several runs on the same input. An overhead thus applies when performing a polynomial product where two direct TFT and one inverse TFT have to be performed. Taking also into account zero padding and the entry-wise vector product in between, the total time sensibly exceeds three times the one displayed in Table 11.

| $d$ | $2^8$ | $2^9$ | $2^{10}$ | $2^{11}$ | $2^{12}$ | $2^{13}$ | $2^{14}$ | $2^{15}$ | $2^{16}$ | $2^{17}$ | $2^{18}$ | $2^{19}$ | $2^{20}$ |
|---|---|---|---|---|---|---|---|---|---|---|---|---|---|
| NTL | 0.059 | 0.18 | 0.24 | 0.52 | 1.1 | 2.4 | 5.2 | 11 | 23 | 49 | 100 | 250 | 580 |
| FLINT | 0.017 | 0.043 | 0.11 | 0.29 | 0.71 | 1.8 | 4.6 | 11 | 22 | 48 | 110 | 290 | 610 |
| Kronecker | 0.031 | 0.076 | 0.19 | 0.46 | 1.2 | 2.6 | 5.8 | 12 | 30 | 60 | 140 | 310 | 680 |
| Math. | 0.0068 | 0.013 | 0.026 | 0.054 | 0.11 | 0.24 | 0.51 | 1.1 | 2.5 | 5.4 | 12 | 26 | 60 |

**Table 14.** Polynomial product for degrees $<d$ over $\mathbb{Z}/469762049\,\mathbb{Z}$, user time in milliseconds.

If $K$ is a field with sufficiently many primitive roots of unity of order a power of two, then two $n \times n$ matrices $A$ and $B$ with coefficients in $K[x]$ of degrees $<d$ can be multiplied by performing TFT transforms of length $2\,d$ on each of the coefficients of $A$ and $B$, by multiplying $2\,d$ matrices over $K$, and finally by recovering the matrix product through coefficient-wise inverse transforms. This requires $O(d\,n^\omega + n^2\,d\,\log d)$ operations in $K$, where $\omega \leqslant 3$ is the exponent of $n \times n$ matrix multiplication over $K$. In general, most of the time is spent in the matrix products over $K$. Nevertheless, if $n$ remains sufficiently small with respect to $d$, then most of the time is spent in the fast Fourier transforms, and our method becomes most efficient. Table 15 compares this approach to naive multiplication, and to FLINT.

| $n$ | 1 | 2 | 4 | 8 | 16 | 32 |
|---|---|---|---|---|---|---|
| Naive | 0.0012 | 0.098 | 0.079 | 0.65 | 5.2 | 40 |
| FLINT | 0.011 | 0.086 | 0.70 | 5.6 | 45 | 361 |
| Mathemagix | 0.0012 | 0.0067 | 0.031 | 0.15 | 0.80 | 4.7 |

**Table 15.** Polynomial matrix product over $\mathbb{Z}/469762049\,\mathbb{Z}$ for degrees $d < 2^{15}$, user time in seconds.

## 5.3. Integer matrix product

Another important application of fast Fourier transforms is integer multiplication. The method that we have implemented is based on *Kronecker segmentation* and the *three-prime FFT* (see [55] and [26, Section 8.3]). Let $p_1$, $p_2$ and $p_3$ be three prime numbers. The two integers $a$ and $b$ of at most $N$ bits to be multiplied are split into chunks of a suitable bit-size $h$ and converted into polynomials $A$ and $B$ of $\mathbb{Z}[X]$ of degrees $<d$, with $A(2^h) = a$, $B(2^h) = b$ and $d = \lceil N/h \rceil$. The maximum bit-size of the coefficients of the product $C(X) = A(X)\,B(X)$ is at most $H = 2\,h + \lceil \log(d+1)/\log 2 \rceil$. The parameter $h$ is taken such that $d$ is minimal under the constraint that $2^H < p_1\,p_2\,p_3$. The polynomial $C$ can then be recovered from its values computed modulo $p_1$, $p_2$ and $p_3$ using TFT multiplications. Table 16 shows the performance obtained by this approach for $p_1 = 998244353$, $p_2 = 985661441$ and $p_3 = 943718401$. We compare with the timings of GMP.

| $n$ | $2^8$ | $2^9$ | $2^{10}$ | $2^{11}$ | $2^{12}$ | $2^{13}$ | $2^{14}$ | $2^{15}$ | $2^{16}$ | $2^{17}$ | $2^{18}$ | $2^{19}$ | $2^{20}$ |
|---|---|---|---|---|---|---|---|---|---|---|---|---|---|
| GMP | 0.0060 | 0.017 | 0.045 | 0.12 | 0.32 | 0.81 | 1.9 | 4.2 | 9.7 | 21 | 48 | 110 | 240 |
| Mathemagix | 0.028 | 0.053 | 0.11 | 0.21 | 0.44 | 0.92 | 2.0 | 4.6 | 10 | 21 | 46 | 91 | 200 |

**Table 16.** Integer product in bit-size $32 \times 2^n$, user time in milliseconds.

Similarly to polynomial matrices, small matrices over large integers can be multiplied efficiently using modular TFTs. In Table 17, the row "Mathemagix" shows timings for this approach. The row "Naive" corresponds to the classical product calling GMP functions directly. We also compare to FLINT and LinBox version 1.3.2 [18] (using NTL and FFlas-FFpack 1.6.0 [19, 20]).

| $n$ | 1 | 2 | 4 | 8 | 16 | 32 |
|---|---|---|---|---|---|---|
| Naive | 0.0042 | 0.033 | 0.27 | 2.2 | 17 | 140 |
| LinBox | 0.0042 | 0.033 | 0.27 | 2.2 | 17 | 140 |
| Flint | 0.0042 | 0.036 | 0.29 | 2.3 | 19 | 150 |
| Mathemagix | 0.0060 | 0.022 | 0.095 | 0.44 | 2.3 | 13 |

**Table 17.** Integer matrix product in bit-size $32 \times 2^{15}$, user time in seconds.

## Conclusion

Nowadays SIMD technology has widely proved to be extremely efficient for numerical computations. A major conclusion of the present work is that modular integers, and therefore exact and symbolic computations can also greatly benefit from this technology *via* integer types, and even in small sizes. We believe that the dissemination of the recent AVX-512 and other forthcoming extensions with wider vectors will be very profitable.

The use of C++ as a programming language allowed us to develop generic template libraries inside Mathemagix, which are both efficient and uniform. Nevertheless, ensuring good performance for the implementations presented here was not an easy task. Depending on the presence or absence of specific SIMD instructions, we had to implement a large number of variants for the same algorithms. Template programming turns out to be essential to control loop unrolling and memory alignment. Furthermore, the efficiency of several elementary routines for small fixed sizes could only be ensured through hand optimization of very specific pieces of code. For instance, several FFTs in small fixed sizes, say the size of SIMD registers, cannot be easily vectorized optimally. Here hardware transposition routines would be helpful, for example for $8 \times 8$ matrices of 32-bit integers using AVX 2 registers. On the longer run, we expect that part of the work that we did by hand could be done by an optimizing compiler. One major challenge for compilers and the design of programming languages is the automatic generation of FFTW3-style codelets.

As a final comment we would like to emphasize that SIMD technology can be seen as a low level way to parallelize computations. We believe that bringing easy access to this technology in Mathemagix will be fruitful to non specialist computer algebra developers. Currently, such developers are faced to monolithic software such as Maple™ [28] or Magma [14], which very well cover low and high user-level mathematical functions, but lack of general lower level programming facilities. Some recent C++ and Python projects [56] aim at providing frameworks for large scale scientific programming, such as [13, 59], but all gaps have not been filled yet. One example of an interesting practical challenge is the multiplication of dense matrices with large integer coefficients. Indeed, this was one of the programming contest of the PASCO 2010 conference, and our benchmarks in Section 5.3 show that there was indeed room for progress.

## Bibliography

[1] D. Abrahams and A. Gurtovoy. *C++ Template Metaprogramming: Concepts, Tools, and Techniques from Boost and Beyond*. Addison Wesley, 2004.

[2] A. V. Aho, J. E. Hopcroft, and J. D. Ullman. *The design and analysis of computer algorithms*. Addison-Wesley series in computer science and information processing. Addison-Wesley Pub. Co., 1974.

[3] R. Alverson. Integer division using reciprocals. In *Proceedings of the Tenth Symposium on Computer Arithmetic*, pages 186–190. IEEE Computer Society Press, 1991.

[4] H. G. Baker. Computing A*B (mod N) efficiently in ANSI C. *SIGPLAN Not.*, 27(1):95–98, 1992.

[5] B. Bank, M. Giusti, J. Heintz, G. Lecerf, G. Matera, and P. Solernó. Degeneracy loci and polynomial equation solving. Accepted for publication to Foundations of Computational Mathematics. Preprint available from http://arxiv.org/abs/1306.3390, 2013.

[6] N. Bardis, A. Drigas, A. Markovskyy, and J. Vrettaros. Accelerated modular multiplication algorithm of large word length numbers with a fixed module. In M. D. Lytras, P. Ordonez de Pablos, A. Ziderman, A. Roulstone, H. Maurer, and J. B. Imber, editors, *Organizational, Business, and Technological Aspects of the Knowledge Society*, volume 112 of *Communications in Computer and Information Science*, pages 497–505. Springer Berlin Heidelberg, 2010.

[7] P. Barrett. Implementing the Rivest Shamir and Adleman public key encryption algorithm on a standard digital signal processor. In A. Odlyzko, editor, *Advances in Cryptology – CRYPTO' 86*, volume 263 of *Lect. Notes Comput. Sci.*, pages 311–323. Springer Berlin Heidelberg, 1987.

[8]  D. J. Bernstein, Hsueh-Chung Chen, Ming-Shing Chen, Chen-Mou Cheng, Chun-Hung Hsiao, Tanja Lange, Zong-Cing Lin, and Bo-Yin Yang. The billion-mulmod-per-second PC. In *SHARCS'09 Special-purpose Hardware for Attacking Cryptographic Systems: 131*, 2009. http://cr.yp.to/djb.html.

[9]  D. J. Bernstein, Tien-Ren Chen, Chen-Mou Cheng, Tanja Lange, and Bo-Yin Yang. ECM on graphics cards. In A. Joux, editor, *Advances in Cryptology - EUROCRYPT 2009*, volume 5479 of *Lect. Notes Comput. Sci.*, pages 483–501. Springer Berlin Heidelberg, 2009.

[10] J. Berthomieu, G. Lecerf, and G. Quintin. Polynomial root finding over local rings and application to error correcting codes. *Appl. Alg. Eng. Comm. Comp.*, 24(6):413–443, 2013.

[11] J. Berthomieu, J. van der Hoeven, and G. Lecerf. Relaxed algorithms for $p$-adic numbers. *J. Théor. Nombres Bordeaux*, 23(3):541–577, 2011.

[12] D. Bini and V. Y. Pan. *Polynomial and Matrix Computations: Fundamental Algorithms*. Progress in Theoretical Computer Science. Birkhauser Verlag GmbH, 2012.

[13] Boost team. Boost (C++ libraries). Software available at http://www.boost.org, from 1999.

[14] W. Bosma, J. Cannon, and C. Playoust. The Magma algebra system. I. The user language. *J. Symbolic Comput.*, 24(3-4):235–265, 1997.

[15] A. Bosselaers, R. Govaerts, and J. Vandewalle. Comparison of three modular reduction functions. In D. R. Stinson, editor, *Advances in Cryptology — CRYPTO' 93*, volume 773 of *Lect. Notes Comput. Sci.*, pages 175–186. Springer Berlin Heidelberg, 1994.

[16] British Standards Institution. *The C standard: incorporating Technical Corrigendum 1: BS ISO/IEC 9899/1999*. John Wiley, 2003.

[17] CLANG, a C language family frontend for LLVM. Software available at http://clang.llvm.org, from 2007.

[18] J.-G. Dumas, T. Gautier, C. Pernet, and B. D. Saunders. LinBox founding scope allocation, parallel building blocks, and separate compilation. In K. Fukuda, J. van der Hoeven, M. Joswig, and N. Takayama, editors, *Mathematical Software – ICMS 2010*, volume 6327 of *Lect. Notes Comput. Sci.*, pages 77–83. Springer Berlin Heidelberg, 2010.

[19] J.-G. Dumas, P. Giorgi, and C. Pernet. FFPACK: Finite field linear algebra package. In J. Schicho, editor, *Proceedings of the 2004 International Symposium on Symbolic and Algebraic Computation*, ISSAC '04, pages 119–126. ACM Press, 2004.

[20] J.-G. Dumas, P. Giorgi, and C. Pernet. Dense linear algebra over word-size prime fields: The FFLAS and FFPACK packages. *ACM Trans. Math. Softw.*, 35(3):19:1–19:42, 2008.

[21] A. Fog. *Instruction tables. Lists of instruction latencies, throughputs and micro-operation breakdowns for Intel, AMD and VIA CPUs*. http://www.agner.org/optimize, Copenhagen University College of Engineering, 2012.

[22] A. Fog. *Optimizing software in C++. An optimization guide for Windows, Linux and Mac platforms*. http://www.agner.org/optimize, Copenhagen University College of Engineering, 2012.

[23] A. Fog. *Optimizing subroutines in assembly language. An optimization guide for x86 platforms*. http://www.agner.org/optimize, Copenhagen University College of Engineering, 2012.

[24] L. Fousse, G. Hanrot, V. Lefèvre, P. Pélissier, and P. Zimmermann. MPFR: A multiple-precision binary floating-point library with correct rounding. *ACM Trans. Math. Software*, 33(2), 2007. Software available at http://www.mpfr.org.

[25] M. Frigo and S. G. Johnson. The design and implementation of FFTW3. *Proc. IEEE*, 93(2):216–231, 2005.

[26] J. von zur Gathen and J. Gerhard. *Modern computer algebra*. Cambridge Univ. Press, 2nd edition, 2003.

[27] GCC, the GNU Compiler Collection. Software available at http://gcc.gnu.org, from 1987.

[28] K. Geddes, G. Gonnet, and Maplesoft. Maple (TM). http://www.maplesoft.com/products/maple, from 1980.

[29] P. Giorgi, Th. Izard, and A. Tisserand. Comparison of modular arithmetic algorithms on GPUs. In B. Chapman, F. Desprez, G. R. Joubert, A. Lichnewsky, F. Peters, and Th. Priol, editors, *Parallel Computing: From Multicores and GPU's to Petascale*, volume 19 of *Advances in Parallel Computing*, pages 315–322. IOS Press, 2010.

[30] P. Giorgi and R. Lebreton. Online order basis algorithm and its impact on block Wiedemann algorithm. In *Proceedings of the 2014 International Symposium on Symbolic and Algebraic Computation*, ISSAC '14. ACM Press, 2014. To appear.

[31] T. Granlund et al. GMP, the GNU multiple precision arithmetic library, from 1991. Software available at http://gmplib.org.

[32] T. Granlund and P. L. Montgomery. Division by invariant integers using multiplication. In *Proceedings of the ACM SIGPLAN 1994 conference on Programming language design and implementation*, PLDI '94, pages 61–72, New York, NY, USA, 1994. ACM Press.

**[33]** Sardar Anisul Haque and Marc Moreno Maza. Plain polynomial arithmetic on GPU. *Journal of Physics: Conference Series*, 385(1):012014, 2012.

**[34]** W. Hart and the FLINT Team. FLINT: Fast Library for Number Theory, from 2008. Software available at http://www.flintlib.org.

**[35]** W. Hart and the MPIR Team. MPIR, Multiple Precision Integers and Rationals, from 2010. Software available at http://www.mpir.org.

**[36]** D. Harvey. A cache-friendly truncated FFT. *Theoret. Comput. Sci.*, 410(27–29):2649–2658, 2009.

**[37]** D. Harvey and D. S. Roche. An in-place truncated Fourier transform and applications to polynomial multiplication. In S. M. Watt, editor, *Proceedings of the 2010 International Symposium on Symbolic and Algebraic Computation*, ISSAC '10, pages 325–329, New York, NY, USA, 2010. ACM Press.

**[38]** D. Harvey and A. V. Sutherland. Computing Hasse–Witt matrices of hyperelliptic curves in average polynomial time. *Algorithmic Number Theory 11th International Symposium (ANTS XI)*, 2014. To appear.

**[39]** W. Hasenplaugh, G. Gaubatz, and V. Gopal. Fast modular reduction. In P. Kornerup and J.-M. Muller, editors, *18th IEEE Symposium on Computer Arithmetic, ARITH '07*, pages 225–229. IEEE Computer Society, 2007.

**[40]** J. van der Hoeven. The truncated Fourier transform and applications. In J. Schicho, editor, *Proceedings of the 2004 International Symposium on Symbolic and Algebraic Computation*, ISSAC '04, pages 290–296. ACM Press, 2004.

**[41]** J. van der Hoeven and G. Lecerf. Interfacing Mathemagix with C++. In M. Monagan, G. Cooperman, and M. Giesbrecht, editors, *Proceedings of the 2013 International Symposium on Symbolic and Algebraic Computation*, ISSAC '13, pages 363–370. ACM Press, 2013.

**[42]** J. van der Hoeven and G. Lecerf. *Mathemagix User Guide*. HAL, 2013. http://hal.archives-ouvertes.fr/hal-00785549.

**[43]** J. van der Hoeven and G. Lecerf. On the bit-complexity of sparse polynomial and series multiplication. *J. Symbolic Comput.*, 50:227–254, 2013.

**[44]** J. van der Hoeven, G. Lecerf, B. Mourain, Ph. Trébuchet, J. Berthomieu, D. Diatta, and A. Mantzaflaris. Mathemagix, the quest of modularity and efficiency for symbolic and certified numeric computation. *ACM SIGSAM Communications in Computer Algebra*, 177(3), 2011. In Section "ISSAC 2011 Software Demonstrations", edited by M. Stillman, p. 166–188.

**[45]** J. van der Hoeven, G. Lecerf, B. Mourrain, et al. Mathemagix, 2002–2014. Software available at http://www.mathemagix.org.

**[46]** Intel Corporation. Intel (R) intrinsics guide. Version 3.0.1, released 7/23/2013. http://software.intel.com/en-us/articles/intel-intrinsics-guide.

**[47]** Intel Corporation, 2200 Mission College Blvd., Santa Clara, CA 95052-8119, USA. *Intel (R) Architecture Instruction Set Extensions Programming Reference*, 2013. Ref. 319433-015. http://software.intel.com/en-us/intel-isa-extensions.

**[48]** Ç. Kaya Koç. Montgomery reduction with even modulus. *IEE Proceedings - Computers and Digital Techniques*, 141(5):314–316, 1994.

**[49]** Ç. Kaya Koç, T. Acar, and Jr. Kaliski, B. S. Analyzing and comparing Montgomery multiplication algorithms. *Micro, IEEE*, 16(3):26–33, 1996.

**[50]** D. E. Knuth. *The Art of Computer Programming, Volume 2: Seminumerical Algorithms*. Pearson Education, 3rd edition, 1997.

**[51]** G. Lecerf. Mathemagix: towards large scale programming for symbolic and certified numeric computations. In K. Fukuda, J. van der Hoeven, M. Joswig, and N. Takayama, editors, *Mathematical Software - ICMS 2010, Third International Congress on Mathematical Software, Kobe, Japan, September 13-17, 2010*, volume 6327 of *Lect. Notes Comput. Sci.*, pages 329–332. Springer, 2010.

**[52]** P. L. Montgomery. Modular multiplication without trial division. *Math. Comp.*, 44(170):519–521, 1985.

**[53]** M. Moreno Maza and Y. Xie. FFT-Based Dense Polynomial Arithmetic on Multi-cores. In D. J. K. Mewhort, N. M. Cann, G. W. Slater, and T. J. Naughton, editors, *High Performance Computing Systems and Applications*, volume 5976 of *Lect. Notes Comput. Sci.*, pages 378–399. Springer Berlin Heidelberg, 2010.

**[54]** N. Nedjah and L. de Macedo Mourelle. A review of modular multiplication methods and respective hardware implementations. *Informatica*, 30(1):111–129, 2006.

**[55]** J. M. Pollard. The fast Fourier transform in a finite field. *Math. Comp.*, 25(114):365–374, 1971.

**[56]** G. van Rossum and J. de Boer. Interactively testing remote servers using the Python programming language. *CWI Quarterly*, 4(4):283–303, 1991. Software available at http://www.python.org.

**[57]** M. J. Schulte, J. Omar, and E. E. Jr. Swartzlander. Optimal initial approximations for the Newton-Raphson division algorithm. *Computing*, 53(3-4):233–242, 1994.

**[58]** V. Shoup. *NTL: A Library for doing Number Theory*, 2014. Software, version 6.1.0. http://www.shoup.net/ntl.

**[59]** W. A. Stein et al. *Sage Mathematics Software*. The Sage Development Team, from 2004. Software available at http://www.sagemath.org.

**[60]** E. Thomé. Théorie algorithmique des nombres et applications à la cryptanalyse de primitives cryptographiques. http://www.loria.fr/~thome/files/hdr.pdf, 2012. Mémoire d'habilitation à diriger des recherches de l'Université de Lorraine, France.